# Bridging the gap: a comparative study of academic and developer approaches to smart contract vulnerabilities

**Francesco Salzano[1] · Lodovica Marchesi[2] · Cosmo Kevin Antenucci[1] · Simone Scalabrino[1] · Roberto Tonelli[2] · Rocco Oliveto[1] · Remo Pareschi[1]**

## Abstract

In this paper, we investigate the strategies adopted by Solidity developers to fix security vulnerabilities in smart contracts. Vulnerabilities are categorized using the DASP TOP 10 taxonomy, and fixing strategies are extracted from 364 commits collected from open-source Solidity projects on GitHub. Each commit was selected through a two-phase process: an initial filter using natural language processing techniques, followed by manual validation. We assessed whether these fixes adhere to established academic guidelines. Our analysis shows that 60.55% of the commits aligned with at least one literature-based recommendation, particularly for well-documented vulnerability types such as Reentrancy and Arithmetic. However, adherence dropped significantly for categories like Denial of Service, Time Manipulation, and Bad Randomness, highlighting gaps between academic best practices and real-world developer behavior. From the remaining 143 non-aligned commits, we identified 27 novel fixing strategies not previously discussed in the literature. To evaluate their quality, we conducted a structured questionnaire involving 9 experts from both academia and industry. Their feedback indicated high perceived effectiveness of the new fixes, especially for vulnerabilities like Reentrancy and Unchecked Return Values. Generalizability received more varied responses, suggesting context-specific applicability. Finally, we performed a post-fix evolution analysis on over 6700 subsequent commits to assess the long-term stability of the fixes. Most patches remained unchanged, confirming their persistence in production code. Our findings offer practical insights into how vulnerabilities are fixed in smart contracts today, reveal promising emerging patterns, and help bridge the gap between academic guidelines and developer practices.

---

---

Extended author information available on the last page of the article

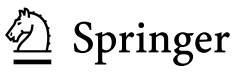

## 1 Introduction

Blockchain technology has garnered significant attention since the introduction of Bitcoin (Nakamoto 2008). Smart Contracts (SCs) are programs that run logic on blockchains, and have seen increasing adoption, becoming responsible for managing high stakes (Zou et al. 2019). Vulnerabilities in the context of blockchain refer to flaws or weaknesses in the design, implementation, or use of blockchain technologies that can be exploited to perform malicious or unwanted actions. These vulnerabilities also exist in the Smart Contracts code. Such vulnerabilities can lead to substantial value losses, as seen in the case of the Decentralized autonomous organization (DAO) attack, which resulted in the malicious withdrawal of cryptocurrencies worth approximately $60 million (Porru et al. 2017).

Therefore, security is crucial, and as a consequence, several vulnerability detection tools have been developed and are available in the literature (Feist et al. 2019; Tikhomirov et al. 2018; Ferreira et al. 2020). In addition, there are empirical studies on their effectiveness (Durieux et al. 2020; Ghaleb and Pattabiraman 2020). These studies focused on Ethereum, in which Solidity serves as a predominant language (Rameder et al. 2022), and we follow the same setting in our work, although Ethereum supports another language for SC development, namely Vyper. Research in the field of software reliability has identified security smells as indicators that may signal underlying security vulnerabilities which can adversely affect the execution and reliability of SCs (Demir et al. 2019a). These security smells serve as early warning signs, alerting developers to potential issues that could lead to significant security breaches if not addressed promptly. Security defects, as defined in current literature, refer to errors that result in incorrect outputs or operational failures within the software (Chen et al. 2020). Such defects encompass a wide range of issues, including both software bugs (meaning programming mistakes that produce unintended behavior) and vulnerabilities, which are specific flaws that could be exploited by attackers to compromise the integrity or confidentiality of the system. Moreover, the literature provides security code recommendations, which are established best practices and guidelines specifically crafted to enhance the security of software code. These recommendations aim to help developers implement more secure coding practices, thereby minimizing the risk of introducing security defects into their applications (Zhou et al. 2023b).

Despite the availability of these guidelines, a significant gap remains in understanding the extent to which developers adhere to them. It is currently unclear whether developers are consistently following the provided fixing strategies or if they are employing alternative, potentially effective strategies when addressing security issues in their smart contracts. This raises important questions about the practices and decision-making processes of Solidity developers in vulnerability-fixing activities.

In this research, we propose a study to bridge these gaps. As SCs are still in the early stages of development, it is important to periodically review security guidelines. Developers may introduce new solutions that can improve existing approaches, and our objective is to analyze these to determine their validity. Our approach involved gathering vulnerability fix recommendations from the existing literature. Subsequently, we examined Solidity GitHub repositories to identify commits addressing vulnerabilities and verify whether these fixes align with the recommendations in the literature. For each type of vulnerability included in the Decentralized Application Security Project (DASP) taxonomy, we reported the level of adherence to literature guidelines in terms of percentage.

Additionally, we collected and analyzed vulnerability fixes that are not covered in the existing literature to assess their suitability for the community. Our findings indicate that only the most documented vulnerabilities receive significant consideration when developers fix their SCs. Conversely, when dealing with several less-studied vulnerability classes in the context of SCs, such as *denial of service* and *time manipulation*, our results show that developers do not follow academic recommendations. This underscores the need to modernize the existing set of fixing approaches. To address this requirement, our study provides new fixing strategies extracted from the commits we analyzed. Specifically, we identified 143 commits containing vulnerability resolution patterns not tracked in the current academic literature, from which we extracted 35 undocumented fixing strategies with 27 distinct approaches along with descriptions that report the underlying motivations supporting their generalizable usage.

To add depth to our mining study, we also evaluated the stability of the gathered fixing commit over time, such an evaluation highlighted the stability of the fixes. The remainder of the paper is organized as follows: Section 2 presents the knowledge required to understand our study; Section 3 presents an overview of the current state of the art; Section 4 summarizes the guidelines collected by reviewing the literature to fix SC vulnerabilities; Section 5 outlines the design of the empirical study we conducted; Section 6 underscores and discusses the achieved findings; Section 7 presents the evaluation of the new fixes and the evaluation regarding the stability over time of the changes made by the collected commits; Section 8 discusses the empirical study results as well as practical development behaviors; Section 9 shows the threats to validity of our work and Section 10 concludes the paper.

## 2 Background

In this Section, we explain the technologies and the concepts involved in our study to ensure understanding of our work.

**Blockchain** Satoshi Nakamoto introduced Blockchain technology as a peer-to-peer cash system in 2008 (Nakamoto 2008). Since then, this technology has expanded beyond the financial sector into many other fields. One significant factor driving its increasing adoption has been the support for Smart Contracts, first enabled by Buterin with the introduction of Ethereum, currently the second most important and widest Blockchain network (Buterin et al. 2013).

The Blockchain is a self-governed peer-to-peer network transaction system that allows secure operation execution, eliminating the need for a trusted third party (Alsunaidi et al. 2019). Transactions are executed on a decentralized ledger composed of linked sequential blocks, with an immutable connection to the predecessor, ensuring the integrity of the chain. Each block stores validated transactions according to a consensus algorithm. The ledger is shared and replicated, and participants in the network can read and write data on it, granting transparent access to its stored data to every network participant.

Blockchain networks are not all alike; instead, they can vary significantly while still adhering to the same basic principles. The main differentiation among blockchain systems lies in managing access permissions to the network's ledger, which can be public or restricted. Access to the ledger divides blockchains into two main categories: permissionless and permissioned.

**Smart Contracts** The concept of SC was introduced in the 1990s by Szabo, initially described as computerized protocols that executed in transactions the terms of a contract (Szabo 1997). Contemporary interpretations consider SCs as event-driven software replicated on decentralized nodes in equal copies, which are set to automatically execute code when certain conditions are met (Zou et al. 2019). Blockchains are immutable, as well as SCs. Although SCs can be made updatable by using a proxy that routes calls to a new implementation, the original contract remains published on the blockchain, maintaining its immutability (Bodell III et al. 2023).

Users or other SCs can interact with SCs by calling them via transactions. Nodes in the Blockchain network validate the transactions; when a transaction is valid, the result of the execution of the logic codified in the SC is written on their local copy of the Blockchain. To reach inclusion in a block, all the nodes must execute this logic in the same way; now stored data are irreversible due to the immutability of the Blockchain. This implies that if a transaction finishes unexpectedly, the result may not be reversible.

**Ethereum & Solidity** Ethereum is the largest blockchain-based smart contract platform, while Bitcoin is the largest cryptocurrency platform. Ethereum enables smart contract execution through the Ethereum Virtual Machine, making it the most widely used smart contract environment. Smart contracts written in high-level programming languages are compiled into Ethereum bytecode, with Solidity being the predominant language used on the Ethereum platform (Zou et al. 2019; Buterin et al. 2014).

Solidity is a programming language that shares a syntax similar to JavaScript, which was introduced in 2015. Since then, its grammar has undergone several changes. The language has received numerous new features, while deprecated ones have been removed. These changes have been made to improve the language's safety and usability (Wang et al. 2021). As a domain-specific language (DSL), it is a programming language of limited expressiveness focused on a particular domain, in essence, it serves mainly for SC development (Wöhrer and Zdun 2020).

**Ethereum's Gas** Gas is the unit of measurement used to determine the work done by Ethereum for interactions within the network. SCs are run by miners on their nodes, and they receive a quantity of gas as a reward. Miners can establish the conditions that transactions must meet in order to be accepted and transmitted through the network using Ethereum clients. For instance, they can set the minimum Gas price required to mine a transaction and determine the desired amount of Gas per block when mining a new block (Pierro and Rocha 2019).

Users requesting transactions pay this reward. Every transaction has a *gas limit* that determines the maximum gas cost. If the cost exceeds the limit, the transaction will be reversed, and an exception will be raised (Chen et al. 2020). In addition to paying for gas, users must also have an Ether (ETH) balance in their wallets to cover the transaction fees.

This ETH is deducted from the user's account when the transaction is executed. Without a sufficient balance of ETH, the transaction will not be processed (Buterin et al. 2013).

**Smart Contract Vulnerabilities** The research refers to the DASP[1] TOP 10 SC vulnerabilities for classifying security issues (Durieux et al. 2020; Ferreira et al. 2020; Dia et al. 2021). The vulnerabilities included in the DASP are listed in Table 1, along with a description.

**Table 1** Vulnerabilities included in the DASP TOP 10 taxonomy and comparison with the SWC Registry vulnerabilities

| DASP TOP 10 Category | Description | SWC Registry Equivalent |
|---|---|---|
| Reentrancy | Contracts are able to call other contracts. Reentrancy occurs when the target contract is recursively called by an external contract before completing the update of its state, leading to an inconsistent state. | SWC-107: Reentrancy |
| Access Control | When there is a lack of secure access and proper authorization to functions, it creates opportunities for attackers to gain direct access to private values or functions, potentially compromising sensitive information and system integrity. | SWC-105: Unprotected Ether Withdrawal, SWC-106: Unprotected SELFDESTRUCT Instruction |
| Arithmetic | Math operations are performed on variables with fixed dimensions. Numbers that exceed these dimensions overflow or underflow. When exploited, arithmetic vulnerabilities can lead to incorrect results, compromising reliability. | SWC-101: Integer Overflow and Underflow |
| Unchecked Calls | Solidity provides low-level calls, such as *call()*, in which the error is not propagated and does not revert the current execution. Instead, these calls return a Boolean value set to false. Failing to check this value can lead to undesirable outcomes. | SWC-123: Unchecked Call Return Value |
| Denial of Service | There are several ways that could lead to denial of service, such as maliciously increasing the gas required to compute a function. For example, sending an array with a huge dimension to a function that loops over it. In this case, if gas block limitations are exceeded, transactions will be reverted. | SWC-113: DoS with Failed Call, SWC-128: DoS With Block Gas Limit |
| Bad Randomness | Randomness is difficult to achieve in blockchains due to the need for consensus. The sources of randomness within Solidity are predictable, allowing malicious users to exploit this predictability. | SWC-120: Weak Sources of Randomness from Chain Attributes |
| Front Running | Transactions need to undergo a waiting period before they are added to a block. A potential attacker could potentially view the transaction pool and add another transaction block before the original one. This process could be exploited to reorder transactions in favor of the attacker. | SWC-114: Transaction Order Dependence |
| Time Manipulation | Decisions are often made based on time-related conditions. The current time is typically obtained using "block.timestamp" or "now" instructions. However, this value comes from the miners and can be maliciously manipulated by them. | SWC-116: Block values as a proxy for time |
| Short Address | Solidity pads shorter arguments to 32 bytes. An attacker may manipulate the data sent, making the smart contract read more data than was sent. | SWC-102: Outdated Compiler Version |
| Unknowns | The DASP TOP 10 highlights a category of vulnerabilities that are currently unknown. | N/A |

[1] https://dasp.co/

To provide further insights, we also include a comparison with vulnerabilities from the SWC Registry (Ethereum Developer Community 2020). Both classifications are designed to identify and describe common vulnerabilities in smart contracts, particularly those developed with Solidity for the Ethereum blockchain. Both systems aim to improve smart contract security by educating developers about potential risks and providing guidelines to avoid them. The DASP TOP 10 focuses on 10 main categories that reflect the most serious and well-known security issues, using educational and concise language. On the other hand, the SWC Registry offers a more granular classification and includes more specific vulnerabilities, such as details on unchecked calls, arithmetic overflow issues, and highly technical attacks like buffer overflow. The table provides a high-level comparison that links vulnerabilities between these two taxonomies, helping to strengthen the study's understanding of the overlaps and potential gaps in the categorization. Some categories in DASP, like Unknowns, do not have a direct match in the SWC Registry, as it focuses on known vulnerabilities. In our analysis, we used the SWC Registry classification and IDs to provide a clear and comparable mapping between known vulnerabilities, as its structure facilitates cross-referencing. However, we acknowledge that the SWC Registry has not been actively maintained since 2020 and may be incomplete. Therefore, we cross-checked our mapping with the most recent EEA EthTrust Security Levels specification (Enterprise Ethereum Alliance 2023), which offers updated guidance for smart contract security. For an additional perspective, particularly regarding alignment with software-level taxonomies, we refer the reader to the SWC–CWE mapping resource provided in Ethereum Developer Community (2020).

## 3 Related Work

This section reviews the existing literature related to SCs vulnerabilities and their fixing approaches, in detail, we carried out our literature review on papers resulting from the following query string: `smart contract AND fix AND (vulnerability OR defect OR recommendation).`

We specifically considered only peer-reviewed journal and conference papers written in English and consequently excluded studies where Solidity was not utilized in the SCs. Additionally, we thoroughly searched sources from popular digital libraries such as IEEE Explore, ACM, ScienceDirect, and Springer.

SCs require a thorough security assessment before being deployed. In a survey conducted by Zou et al., it was found that most of the respondents stated that SC development has a higher requirement for code security compared to traditional development (Zou et al. 2019). This is due to the management of digital assets and the irreversible nature of the transactions involved. Academic research has delved heavily into SC security due to its important role in fulfilling research motivations. Indeed, a plethora of vulnerability detection tools have been released and published, assisting the academic and developer communities in seeking security vulnerabilities. Such tools encompass static analysis tools (Feist et al. 2019; Tikhomirov et al. 2018; Ferreira et al. 2020), fuzzing tools (Jiang et al. 2018), as well Machine Learning and Deep Learning-based tools (Shakya et al. 2022; Zhang et al. 2022). These contributions have recently been accompanied by Large Language Model-based vulnerability scanners, such as Gptscan, a new tool that detects logic vulnerabilities in smart contracts by using Large Language Models such as Generative Pre-training Transformer (Sun et al. 2024).

Duriex et al. carried out an empirical review to assess the effectiveness of vulnerability detectors (Durieux et al. 2020), which led to a suggestion for a high false negative rate. In their study, Ghaleb and Pattabiraman assessed the effectiveness of static analysis tools by intentionally introducing security-related bugs (Ghaleb and Pattabiraman 2020). Their findings align with the work of Duriex et al., highlighting the need to improve the detection performance of smart contract vulnerability detection tools.

As a consequence, despite the availability of a wide range of vulnerability detectors, developers still rely on manual detection of vulnerabilities (Ghaleb 2022), remarking a high awareness of security vulnerabilities. A recent study carried out by Chen et al. employed ChatGPT as a security vulnerability detector on the contracts comprised in the curated dataset shared by Durieux et al. (2020); their LLM-based framework achieved a good recall, nonetheless, the low precision problem is yet to be overcome (Chen et al. 2023b). Therefore, given the low reliance on detection tools, providing developers with security smells and vulnerability mitigation approaches is crucial. Several researchers have made contributions to the topic; Demir et al. conducted a comprehensive review of the existing literature in order to identify various vulnerabilities that must be avoided (Demir et al. 2019b). They also created a catalog of security smells to serve as a reference for developers and security professionals. AutoMESC, proposed by Soud et al., introduces a framework for mining and classifying Ethereum SC vulnerabilities and their fixes. It aims to address the lack of open datasets on SC vulnerabilities (Soud et al. 2023). This tool gathers and categorizes SC vulnerabilities and their fixes using seven well-known detection security tools.

Chen et al. have gone deeper, not only providing an extended set of smells but also a wider range of smells, along with defining 20 types of defects in contracts. These defects are categorized according to potential safety, availability, performance, maintenance, and reuse issues. This categorization underscores the importance of security concerns as well (Chen et al. 2020). An important contribution of this study is the valuable solution provided to address such defects, some of which are devoted to addressing security vulnerabilities of Solidity SCs. Recently, Marchesi et al. proposed a structured collection of design patterns and best practices for Ethereum smart contracts, delivering three actionable checklists covering 12 security-critical areas, thus supporting developers in applying secure design principles systematically (Marchesi et al. 2025). Rosa et al. broaden the understanding of SC maintenance by moving beyond defect detection and investigating developers' real-world practices. Through a qualitative analysis of 590 commits from 14 Solidity repositories, they define two taxonomies: one capturing the reasons for maintenance and another detailing modification patterns. Their findings reveal that most changes are devoted to perfective and corrective maintenance, with refactoring and bug fixing representing the majority of cases. Interestingly, they also highlight frequent aesthetic improvements, such as comment and identifier updates, underscoring developers' attention to code readability (Rosa et al. 2025).

The growing body of knowledge and the increasing number of examples regarding fixing approaches have powered automatic program repair (APR) for SCs. Starting from the promising results brought by Yu et al. (2020), more recent studies and tools have further enhanced SC APR. For instance, Nguyen et al. presented *SGUARD*, an approach developed to automatically transform smart contracts so that they are free of 4 common kinds of vulnerabilities (Nguyen et al. 2021), for which they also shared some strategies to fix vulnerabilities. Moreover, Chen et al. proposed *TIPS*, another automated approach to patch SC security vulnerabilities, and provided fixing patterns in their research (Chen et al. 2020).

A novel related work is accomplished by Zhou et al., who created *SmartREP*, a one-line fixing technique for SC repair (Zhou et al. 2023b). As part of their study on software development, the researchers conducted a literature review to identify vulnerabilities commonly encountered in SC development. Based on their findings, they provided 13 code recommendations to address these vulnerabilities. They also paved the way for explicit studies of code changes related to bugs.

*SmartShield*, introduced by Zhang et al. (2020), is an automated bytecode rectification system that addresses three recurring classes of vulnerabilities in Ethereum smart contracts: reentrancy due to state changes after external calls, integer overflows/underflows, and unchecked return values. Rodler et al. proposed *EVMPatch*, a framework for timely and automated patching of Ethereum smart contracts directly at the bytecode level (Rodler et al. 2021). Their evaluation on over 14,000 real-world contracts showed that EVMPatch can effectively mitigate access control and integer overflow vulnerabilities while maintaining functional correctness and incurring negligible gas overhead. Qian et al. conducted a comprehensive survey of smart contract vulnerability detection techniques, classifying vulnerabilities across the Solidity code (Qian et al. 2022), EVM execution, and blockchain dependency layers. They reviewed over 100 works and categorized detection approaches into five main families, namely, formal verification, symbolic execution, fuzzing, intermediate representation, and deep learning.

A recent work in the field of SC code repair recommendation has been steered by Guo et al., who introduced RLRep, a reinforcement learning-based approach for automatically providing repair recommendations for smart contract developers (Guo et al. 2024). They elaborated deeply on repair recommendations, giving a detailed view of fixing patterns in the shape of code snippets. Wang et al. conducted an empirical study on SC bug fixes in real-world Solidity projects, shedding light on bug-fixing through a multi-faceted analysis, considering file type and amount, fix complexity, bug distribution, and fixes of 46 SC projects (Wang et al. 2023). In such a work, they shared insight into bug-fixing effects and implications. Their findings include the types and the number of files involved during bug fixes, fix actions, and complexity, and bug distribution over 14 distinct categories. Moreover, they supplied information regarding how many bugs have been fixed, how many bugs have been newly introduced, and how developers fix bugs in real-world projects.

In summary, prior research on SC security has primarily focused on detecting vulnerabilities (Durieux et al. 2020; Ghaleb 2022; Wang et al. 2024), cataloging security smells and defects (Chen et al. 2023b; Demir et al. 2019b; Marchesi et al. 2025), and proposing automated repair techniques (Yu et al. 2020; Chen et al. 2023a; Huang et al. 2024a). However, comparatively less attention has been given to understanding how developers actually fix vulnerabilities in practice and to what extent these fixes align with academic recommendations. Our study complements this body of work by empirically bridging academic guidelines with real-world developer practices, thereby highlighting gaps in the literature and surfacing new fixing strategies that can inform both research and practice.

## 4 Literature Guidelines

The goal of this section is to define and present the fixing strategies (namely “literature guidelines”) extracted from prior academic work. These guidelines serve as a benchmark for assessing whether real-world fixes in smart contracts align with established best practices. In our study, we define a literature guideline as a fixing strategy that: (i) addresses a vulnerability type included in the DASP TOP 10 taxonomy, and (ii) is accompanied by an example of secure code within a peer-reviewed research paper. Fixes must be explicitly described as addressing security vulnerabilities and must include practical implementation details. We excluded high-level recommendations lacking executable examples without delivering examples of fixing code, as we mentioned in our registered report (Salzano et al. 2024b). We excluded descriptive fixing approaches without practical implementations because they may provide insufficient guidance for developers in the context of SCs. As highlighted by Zhou et al. (2023a), many vulnerabilities can arise from a single line of code. In such cases, high-level recommendations without executable examples may leave room for subtle mistakes in the actual implementation, ultimately failing to mitigate the vulnerability. For this reason, our guidelines required concrete code-level fixes, ensuring that the suggested strategies could be directly applied and verified as effective in addressing the corresponding vulnerability category. For example, while research frequently identified `block.timestamp` as a vector for time manipulation attacks, none of the reviewed papers provided a practical method to address this issue. For example, while research frequently defined this as a vector for attacks (Chen et al. 2020), highlighting that miners can deliberately shift timestamps to influence contract behavior, none of the reviewed papers provided a practical method to address this issue.

The extraction was conducted by three authors, following the completion of the systematic literature review described in our registered report (Salzano et al. 2024b). Each paper was reviewed to identify distinct and actionable fixing strategies. Strategies were then categorized based on the DASP TOP 10 taxonomy to ensure consistency across vulnerability classes. As a selection criterion, we considered not only the presence of secure code, but also an explicit explanation of why the proposed fix mitigates the corresponding vulnerability, thereby ensuring both practical applicability and a clear security rationale.

As a result of our literature review, after checking the papers resulting from our query, 28 papers were selected to search for guidelines. From 11 of those, we gathered academic guidelines that guided our work. After filtering out descriptive-only fixes, 31 guidelines with practical code examples were retained.

Table 2 summarizes the resulting literature guidelines, grouped by vulnerability type. A complete catalog, including references to the source papers, is available in our replication package (Salzano et al. 2024a), accessible at: https://zenodo.org/records/17105939.

## 5 Empirical Study Design

The purpose of the study we propose is to assess whether developers adhere to the current research guidelines when fixing SC security vulnerabilities, and also to identify any valid fixes that are not covered in the existing literature. The study is aimed at researchers who are interested in SC security. The context of the study is based on a dataset of security vulnerability commits that have been fixed in public Solidity SC repositories.

**Table 2** Summary of literature guidelines indicating known fixing approaches categorized by vulnerability type

| Category | Strategies | Source(s) |
|---|---|---|
| Reentrancy | Use `NonReentrant` modifier from OpenZeppelin; use `send()` or `transfer()` instead of `call()`. | (Chen et al. 2020, 2023a; Nguyen et al. 2021; Zhou et al. 2023b) |
| Access Control | Use `OnlyOwner` modifier from OpenZeppelin. | (Chen et al. 2020, 2023a; Zhou et al. 2023b) |
| Arithmetic Issues | Use SafeMath library from OpenZeppelin; use `require` statements to check arithmetic operations. | (Chen et al. 2020, 2023a; Guo et al. 2024; Nguyen et al. 2021; Zhou et al. 2023b) |
| Unchecked Return Values for Low Level Calls | Check the return value of low-level calls with a `require` or an `if`. | (Chen et al. 2020, 2023a; Zhou et al. 2023b) |
| Denial of Service | Avoid using `transfer()` in loop statements. | (Chen et al. 2020) |
| Bad Randomness | Reviewed literature provided some instructions; however, no practical fixes were provided. | (Chen et al. 2020) |
| Front Running | Require the current allowance to match the expected value or be zero before updating it, preventing malicious transactions from exploiting outdated approvals. | (Guo et al. 2024) |
| Time Manipulation | Reviewed literature provided some instructions; however, no practical fixes were provided. | (Chen et al. 2020) |
| Short Addresses | no practical examples for dealing with short addresses, except for avoiding hard-coded addresses by receiving them as an input parameter. | None |

To achieve this, we analyzed a large dataset of commits from public Solidity smart contract repositories on GitHub. Our methodology follows a linear pipeline structured into six main phases: data collection, relevance filtering using NLP, manual labeling and validation, analysis of adherence to literature, identification of new fixes, and expert-based evaluation.

In order to achieve our objective, we will be guided by the following research questions:

- **RQ**$_1$: To what extent do developers adhere to the fixing guidelines provided in the literature?
- **RQ**$_2$: What are the valid fixing approaches beyond those documented in the literature?

### 5.1 Data Collection

The context of our study is a dataset composed of commits addressing security vulnerabilities included in the DASP TOP 10. This taxonomy was chosen because it gained high popularity (Durieux et al. 2020), which may increase the chances of finding the names of its categories in commit messages. Moreover, even if this taxonomy is outdated, it is still used in recent research (Chen et al. 2024). These commits provide pairs of vulnerable and fixed code, offering valuable insights into fixing procedures.

#### 5.1.1 Repository Selection

We focused our study on Solidity SCs by collecting repositories from GitHub. We applied specific filters to ensure quality and relevance. In particular, we included only repositories written in Solidity that had a star count of 10 or more. The star count in GitHub repositories is a metric used to indicate the popularity or appreciation of a repository among the GitHub community. When users find a repository valuable, interesting, or worth revisiting, they can "star" it. The use of the star count is consistent with prior literature (Dabic et al. 2021; Rosa et al. 2018), where it serves as a proxy for popularity and relevance. As in Dabic et al. Dabic et al. (2021), we selected repositories with at least 10 stars to enhance the scalability of the collection while focusing on widely adopted projects.

No minimum file count was required since smart contracts are usually independent. In practice, no single-file repositories passed the star filter. Using GitHub's API, we retrieved all repositories matching these criteria, yielding a total of 5,874 repositories, all of which were included in our analysis.

#### 5.1.2 Commit Mining

Starting from the gathered repositories, we mined commits using *PyDriller* (Spadini et al. 2018), a framework for extracting data from Git repositories.

We included commits that modified at least one file with the .sol extension, corresponding to Solidity source code. At the same time, we excluded merge commits as well as duplicates. A commit $c$ was regarded as a duplicate if there was at least another commit with the same hash and originating from the same repository URL of $c$. These constraints ensured that we focused on relevant, unique commits affecting Solidity files, setting the stage for further filtering based on commit messages.

To further assess the quality of the repositories included in our dataset, we collected additional metadata through the GitHub API. Table 3 reports the average number of commits and contributors per repository, which confirms the maturity and sustained development activity of the analyzed projects.

**Table 3** Mean values for number of commits and contributors per repository in the dataset

| Statistic | Mean Value |
|---|---|
| Number of Commits | 750.65 |
| Number of Contributors | 10.45 |

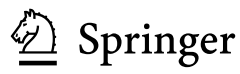

## 5.2 NLP-based Filtering

After mining, to reduce the number of irrelevant commits before manual analysis, we implemented an *NLP-based filter* using *SpaCy*,[2] an open-source NLP library. The pipeline consisted of several steps. First, commit messages were converted to lowercase. They were then *tokenized*, and after that, *stopwords* were removed. Finally, *lemmatization* was applied, for example, transforming a word such as "fixing" into its base form, "fix." These operations were performed using the `en_core_web_lg` SpaCy model. The goal was to standardize messages and retain only their meaningful components.

The filter accepted commits containing the lemma "fix" and lemmas corresponding to DASP TOP 10 vulnerability categories (as shown in Table 1). To tailor the filtering process to security-specific concerns, we explicitly constructed a vocabulary of fixing-related terms (e.g., `fix`, `fixed`, `patch`, `resolve`) and security-specific keywords derived from the DASP Top 10 categories (e.g., `reentrancy`, `access control`, `arithmetic`, `bad randomness`, `denial of service`, `front running`, `time manipulation`, `short address`, and `unchecked low level calls`). To increase recall, we also considered common variants and synonyms such as `overflow`, `underflow`, `race condition`, `timestamp dependence`, or `recursive calls`. The filtering step, therefore, accepted commits whose messages contained both a fixing-related lemma and at least one security-related lemma. For example, a message like *"fix reentrancy vulnerability in token transfer"* would be kept for the analysis by the filter. By grounding the keyword selection in vulnerability categories, the NLP pipeline was specifically oriented toward commits relevant to SC security.

This approach provided a restricted dataset with an adequate number of filtered commits for further analysis. An example of the NLP pipeline output for a real-world commit message is shown in Fig. 1.

The NLP-based filtering process was intentionally designed to accept a high false positive rate in order to minimize the risk of false negatives. This decision was justified by

**Example of input and output of the NLP pipeline tasks on a Commit Message**

**Input Commit Message:**

```
1 fix DOS attack vector for challenging same SB over and over
2 avoid subsequent challenges by checking if a SB has been defended,
3 to not have to have submitter defend it again.
```

**Output:**

```
1 {'dos', 'check', 'defend', 'fix', '.', 'attack', 'sb', 'avoid', ',',
2  'submitter', 'vector', 'subsequent', 'challenge'}
```

**Fig. 1** Example of the NLP pipeline applied to a real-world commit message, showing the extracted lemmatized tokens after processing

[2] https://spacy.io/

the subsequent manual review phase, during which false positives could be discarded. By prioritizing recall over precision at this stage, we ensured that potentially relevant commit messages were not prematurely excluded.

Additionally, heuristics were applied to exclude irrelevant files. Specifically, we discarded files that did not contain a `pragma solidity` declaration, as these are usually not standalone contracts, as well as files ending with the `.t.sol` extension, which are commonly used for testing.

At the end of this filtering process, the candidate dataset consisted of 3,462 Solidity file modifications, derived from the mined commits. It is worth noting that a single commit may affect multiple files.

To increase clarity, we provide a brief data-reduction summary that highlights the stepwise filtering pipeline from the initial commit set to the final dataset used in our analysis. We started from 644,338 raw commits. One repository was excluded because it produced automated updates at the rate of almost one commit per minute, accumulating nearly 3M commits, with a commit message stating *Automated update and the current date*. Therefore, it would not bring relevant data to our dataset, compromising the scalability of the mining phase. After this exclusion, we applied our NLP-based filtering of commit messages, which retained 1,070 commits potentially related to vulnerabilities. Finally, after the manual analysis phase, we kept 364 commits that constitute the dataset for our study. Table 4 summarizes this reduction process.

## 5.3 Manual Labeling and Validation

To assess the relevance and vulnerability category of each commit filtered via the NLP process, we conducted a rigorous manual labeling phase. Each commit was independently evaluated by two of the three authors involved in the labeling process. The evaluators assigned one of the DASP TOP 10 labels (or "Not Relevant") based on the type of vulnerability the commit addressed. When both evaluators agreed on a label, it was directly assigned. In cases of disagreement, all three evaluators participated in a conflict resolution phase.

The inter-rater reliability between evaluators was measured using Cohen's kappa, and we obtained a substantial agreement level of **0.72**. A total of 30 labeling conflicts were identified and resolved by consensus. As part of this phase, we also applied additional filtering criteria. Commits that modified more than three files were excluded, unless the commit message explicitly indicated the vulnerable file or referenced a specific function. Similarly, commits with vague messages were discarded, except in cases where manual inspection allowed a confident identification of the corresponding vulnerability type.

After applying these filters and resolving all conflicts, the final curated dataset consisted of 364 relevant commits assigned to one of the DASP vulnerability categories.

**Table 4** Stepwise data-reduction process from raw commits to the final dataset

| Filtering Step | Remaining Commits |
|---|---|
| Raw commits (all repositories) | 352,535,0 |
| After excluding repository with ∼3M automated commits | 644,338 |
| After NLP-based filtering of commit messages | 1,070 |
| After manual analysis | 364 |

#### 5.3.1 NLP-based Filter Evaluation

At the end of the manual evaluation of the relevance of each commit, 34.02% of the commits that passed the NLP-based filter were confirmed as relevant and assigned to a DASP category.

To evaluate the effectiveness of the NLP-based filtering process, we adopted a sampling-based strategy. Specifically, we randomly selected a set of repositories with a minimum of 50 commits to ensure sufficient development activity. From these, we mined the complete commit history, excluding merge commits, and then randomly sampled 400 commits to construct a statistically meaningful evaluation set.

Each commit in the sample was manually analyzed and annotated with a binary label indicating whether it should have been identified as relevant by the filter. This provided ground truth data for evaluating the system. We then compared this label with the actual output of the filtering system. A commit was marked as a true positive (TP) if it was correctly identified by the filter, or as a false negative (FN) if it was relevant but missed by the filter. Based on this sample, we obtained a recall of 0.8, with 8 true positives identified.

Commits not caught by the filter typically lacked keywords such as “fix” or “vulnerability”, for instance: “add no zero address check when setting beneficiary”. The inclusion of such keywords in the filtering mechanism was a deliberate choice to reduce the number of commits under analysis and focus on those more likely to be related to vulnerability fixes.

The same sample was also used to assess the specificity. The NLP filter correctly ignored all irrelevant commits in the sample, resulting in a measured specificity of 100%. However, this result should be interpreted cautiously. Since the evaluation is based on a sample, the observed specificity is subject to statistical variability. Although false positives (FP) were not observed in the sample, FPs were encountered in the full data set. This discrepancy can be attributed to the confidence interval associated with the estimate: even though specificity in the sample is 100%, the true specificity in the population is likely slightly lower.

This evaluation supports our design choice: by prioritizing high recall, we retained more potentially relevant commits for manual inspection, while avoiding premature exclusions that could have compromised comprehensiveness. Hence, Fig. 2 illustrates the overall workflow carried out to address **RQ**$_1$, summarizing the NLP filtering, manual validation, labeling, and relevance evaluation steps discussed above.

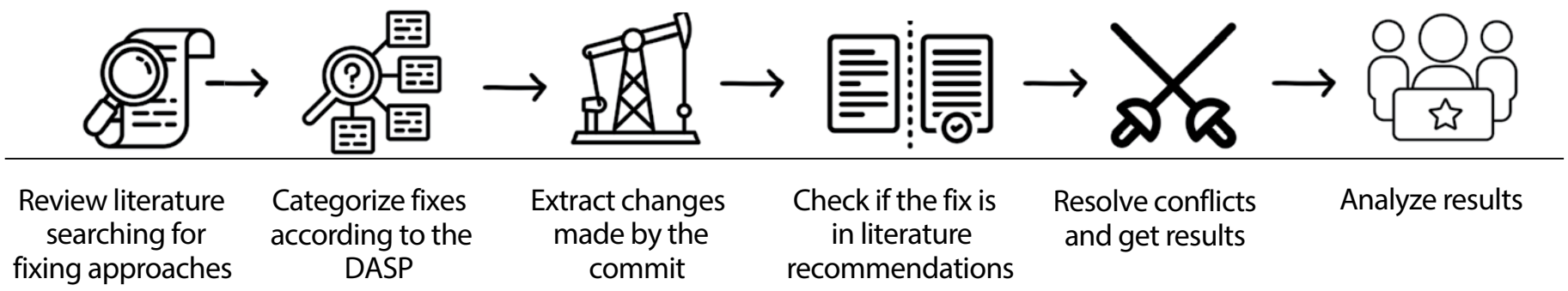

**Fig. 2** Overall workflow to answer **RQ**$_1$

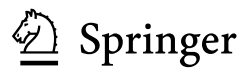

### 5.4 Asserting Adherence to Literature Guidelines

To determine whether developers follow known practices when fixing smart contract vulnerabilities, we compared the code changes made in the manually validated dataset of 364 relevant commits against the set of fixing strategies previously extracted from the literature.

For each commit, we extracted the SC before and after the fixing commit, creating pre-fix and post-fix versions for each commit. Code changes were extracted by comparing the two versions using the *diff_parsed* property of the dictionary returned by PyDriller, which represents a single commit. This dictionary contains two keys: "added" and "deleted," which hold the added and deleted lines, respectively. Originally, we relied on such a diff to see the main differences introduced by a commit, this served particularly in our preliminary analysis to get an initial view of the changes made.

However, the different diff algorithms in Pydriller could influence the results (Nugroho et al. 2020), in order to address this risk and to enhance the readability of these changes, we used a web application capable of showing the difference made by a Git commit, highlighting changes[3] that we integrated into the scripts we use to help us during evaluations. Such a web application displays the entire content for each file modified by a specific commit, highlighting the lines with differences between the version before and after the commit. Evaluators also accessed GitHub's enriched information, such as Pull Request (PR) descriptions and discussions, when additional context was needed. Not all the commits were linked to *PRs*, however, when these were available, we inspected them if we needed additional details to make our decision reliable. Indeed, the description enclosed in the *PRs* guided us to understand the motivation behind several fixes. These two options provided flexibility: the web application offered quick overviews, while GitHub supplied detailed context when required.

A fix was considered a change in which at least one row of the SC containing the vulnerability was modified; differences related to spaces, indentation, and empty rows were ignored. Fixing required actual changes, so if vulnerable lines were simply removed without a replacement, the changes were not considered a fix. When commits included multiple changes, the evaluators identified and isolated the relevant fix from the other changes. The evaluators then determined whether the differences between pre-fix and post-fix versions of the Solidity SCs could be attributed to mitigations available in the literature.

To this end, the two evaluators independently analyzed the commit instances, determining whether each instance contained resolution strategies previously identified in the literature. In cases of conflict, the evaluators discussed their findings until a consensus was reached. When disagreements persisted, a third evaluator reviewed the instance to finalize the decision. Discrepancies were documented, highlighting specific points of contention and outlining differing perspectives on whether the change aligned with literature-reported mitigations.

The inter-rater reliability between the two evaluators was calculated using Cohen's kappa coefficient, which measures the level of agreement (Cohen 1960). This ensured the reliability of the manual analysis and provided a quantitative measure of consistency.

At the end of this step, we provided results showing how many fixes adhere to literature recommendations. To address $\mathbf{RQ}_1$, we report the number and percentage of fixing commits that adopted approaches known in the literature, categorized by each DASP category, and identify the most frequently fixed vulnerabilities. For each category, the computed percentage indicates the extent to which developers adhered to literature fixing guidelines. Fixing approaches not included in the collection of literature guidelines were further analyzed to address $\mathbf{RQ}_2$.

[3] https://diff2html.xyz/

### 5.5 Discovery and Validation of New Fixes

To answer $\mathbf{RQ}_2$, we focused on identifying fixing approaches that were not covered by existing literature guidelines. The overall process is illustrated in Fig. 3. Our starting point was the set of 143 commits that were previously labeled with a DASP vulnerability category (as described in Section 5.5) but did not align with any of the known literature-based fixing strategies. We examined these commits to determine whether they represented valid, novel fixing approaches. The evaluation was conducted by a team of three experts: two researchers with experience in smart contract security and one blockchain practitioner. Each evaluator received a shared set of written evaluation guidelines that defined the criteria for identifying valid fixes.

These criteria were defined to ensure both the practical and conceptual soundness of each fix:

- **Technical Soundness:** Does the change effectively eliminate or mitigate the vulnerability? And does it avoid introducing new risks or unintended behavior in the smart contract?
- **Theoretical Robustness:** Is the fix logically consistent with the attack model? And does it fully remove or reduce the exploitable surface in theory?
- **Long-Term Stability:** Does the fix provide a robust and maintainable solution over time, or could it lead to future compatibility or maintenance issues?
- **Adaptability:** Can the fix be generalized beyond the specific instance in which it was applied, and is it suitable for evolving scenarios typical in smart contract development?
- **Applicability:** Can the fix be reasonably reused in other smart contracts affected by similar vulnerabilities, or is it too context-specific to be broadly useful?

Evaluators collaboratively reviewed each candidate fix and discussed it in depth based on these five dimensions. Rather than assigning numerical scores, the goal was to reach a consensus on whether the fix could be considered a valid, generalizable strategy for addressing the associated DASP vulnerability category. Importantly, we did not assign new labels to the commits in this phase. Each fix retained the DASP category previously assigned during the manual labeling process described in $\mathrm{RQ}_1$. Emerging valid fixes were grouped by DASP category, and recurring patterns were abstracted into generalizable fixing strategies.

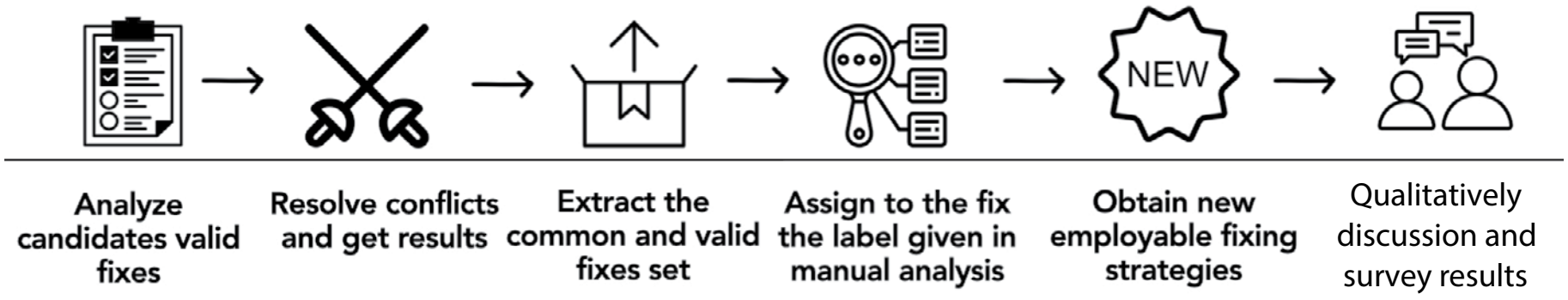

**Fig. 3** Overall workflow to answer $\mathbf{RQ}_2$

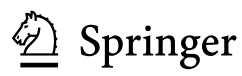

For each of the 27 newly identified strategies, we provide a qualitative description and practical examples in the results section. The full list, including explanations and examples, is available in the replication package.

To further validate such new fixes, these were passed through one more validation phase, made with a survey, which involved 9 experts, as we detail in the next sections of the paper.

### 5.6 Expert-Based Evaluation via Questionnaire

To further validate the newly identified fixing strategies, we designed and administered a questionnaire targeting domain experts. This evaluation method was chosen to collect structured and systematic feedback from both academic and industry professionals with expertise in smart contract security. The goal was to assess each fix across three critical dimensions: generalizability to similar contexts, long-term sustainability, and effectiveness in mitigating the associated vulnerability.

#### 5.6.1 Questionnaire Design

The questionnaire consisted of two main parts: a background section to collect participant information, and an evaluation section focused on the 27 novel fixes.

In the background section, participants were asked to indicate their professional background (academic or industry), years of experience with smart contract development and security, and their familiarity with the DASP TOP 10 vulnerability taxonomy. This information was used to contextualize the collected responses based on the participants' level of expertise.

In the evaluation section, each fix was rated on a 5-point Likert scale (1 = very low, 5 = very high) across the following three dimensions:

- **Generalizability:** How applicable is the fix to similar or recurring cases?
- **Long-Term Sustainability:** Can the fix remain effective and maintainable over time, even as the codebase or context evolves?
- **Effectiveness:** How well does the fix resolve the identified vulnerability?

Each fix was accompanied by a detailed description, a practical code example, and a unique identifier (as introduced in Section 6.2), enabling evaluators to assess the strategies consistently and with adequate context.

The responses collected through this questionnaire are analyzed and discussed in Section 7.1, where we report both aggregated statistics and insights based on the expert feedback.

## 6 Empirical Study Results

In this section, we present the results of our experiments and address the $\mathbf{RQ}_s$ that guided our study.

### 6.1 $RQ_1$: Developer Adherence

Following the methodology described in Section 5, we established a curated dataset of 364 commits, each of which corresponds to a real-world fix for a DASP-classified smart contract vulnerability.

Out of the 364 commits, 221 (60.55%) matched at least one fixing strategy documented in the literature, while 143 introduced novel approaches not previously tracked. This main result highlights that developers often rely on established security practices, but also frequently adopt strategies not yet consolidated in guidelines.

Most vulnerabilities in the dataset relate to arithmetic and reentrancy fixes, while short address attacks and front-running are rarely addressed. Figure 4 shows the distribution of the fixing commits we collected across the DASP TOP 10 vulnerability taxonomy.

To assess adherence to literature guidelines, two raters independently evaluated whether each fixing strategy matched those already documented.

At the end of this analysis, we obtained a Cohen's Kappa value of 0.77, indicating good agreement. Conflicts were resolved with the involvement of a third author to minimize bias. Figure 5 reports the percentage of adherence for each vulnerability class.

We do not report classes with 0% of adherence, namely, DOS, bad randomness, time manipulation, and short address. On the other hand, the front running class showed a moderate adherence level (33.33%), followed by unchecked return values for low level calls (42.86%). Access control, arithmetic, and reentrancy demonstrated high adherence rates, specifically 75%, 66.08%, and 67.95%, respectively.

High adherence percentages suggest that many of the fixing strategies are already well-documented and employed in practice.

To provide more insights, for each vulnerability class, Table 5 details the number of commits matching and not matching guidelines. Specifically, we observe that for Access Control, 51 out of 68 instances matched guidelines, while for Arithmetic, 113 out of 171 matched. Reentrancy shows a similar trend, with 53 out of 78 instances adhering to guide-

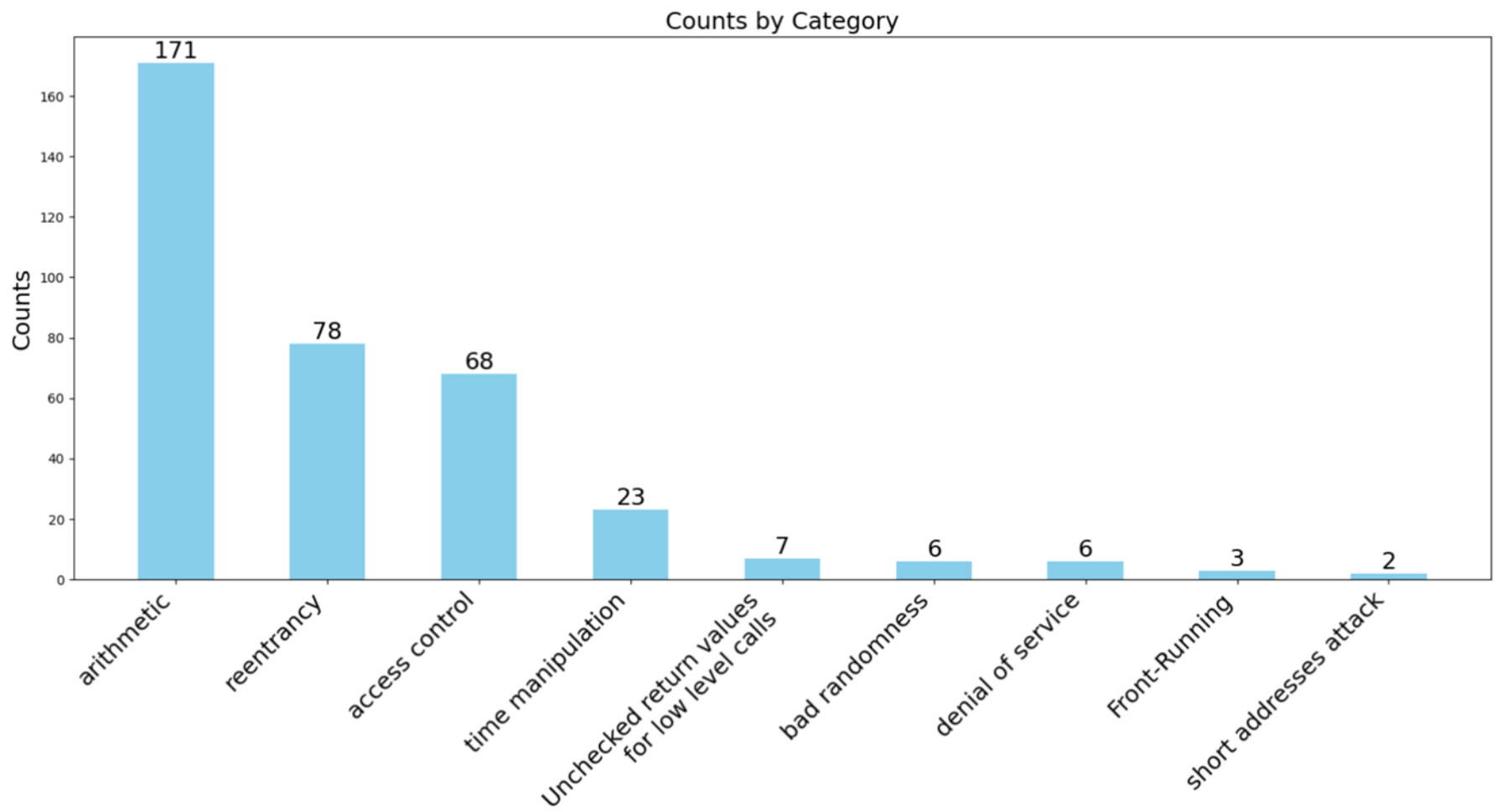

**Fig. 4** Distribution of fixing commits across the DASP TOP 10 vulnerability classes in the analyzed sample

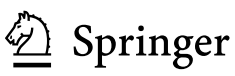

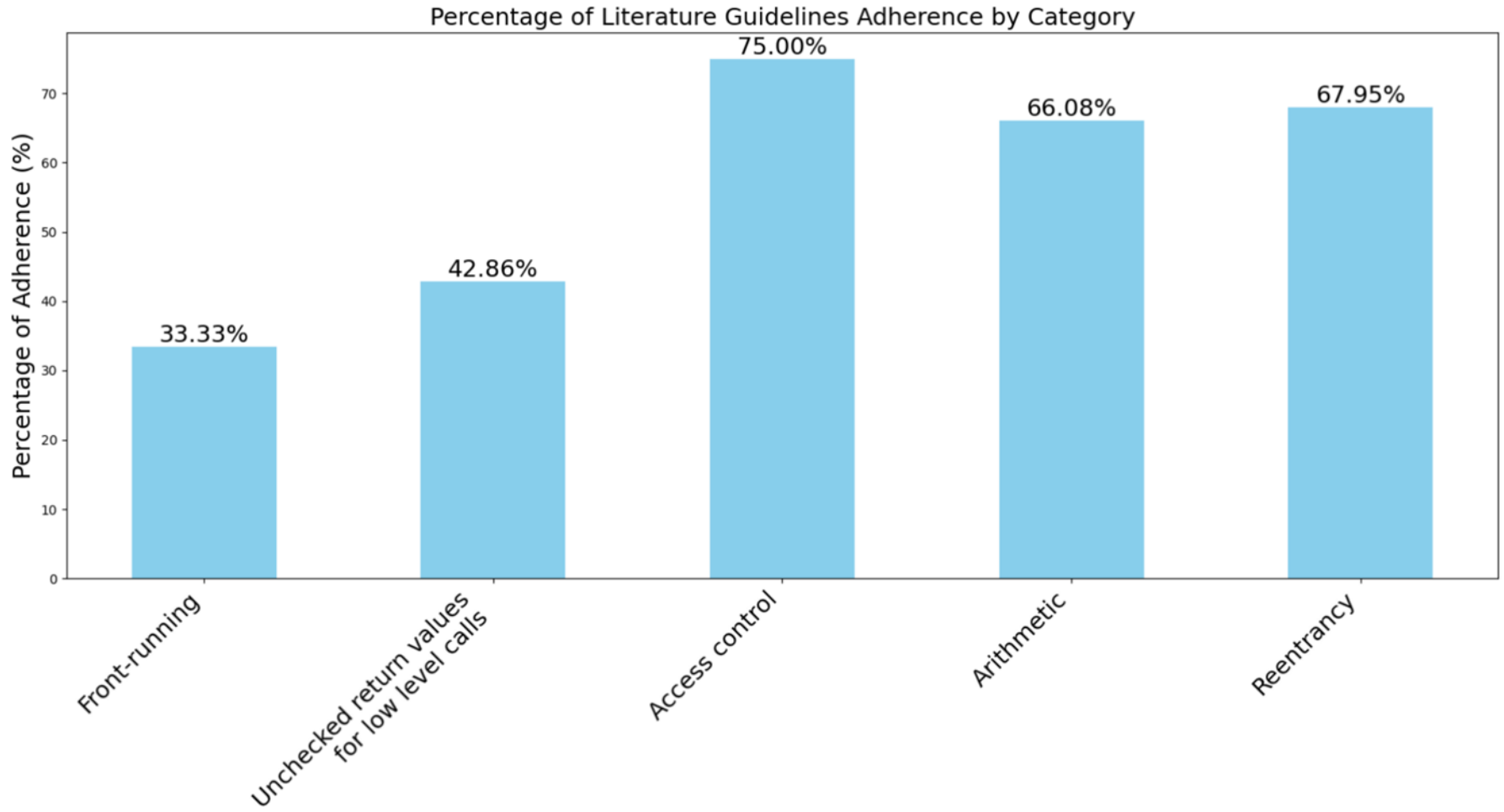

**Fig. 5** Percentage of adherence to literature-documented fixing strategies for each DASP TOP 10 vulnerability category

**Table 5** Distribution of instances that match and do not match Guidelines across Vulnerability Classes

| Vulnerability Class | Matching | Not Matching | Total |
|---|---|---|---|
| Access Control | 51 | 17 | 68 |
| Arithmetic | 113 | 58 | 171 |
| Bad Randomness | 0 | 6 | 6 |
| DOS | 0 | 6 | 6 |
| Front Running | 1 | 2 | 3 |
| Reentrancy | 53 | 25 | 78 |
| Short Address | 1 | 2 | 3 |
| Time Manipulation | 0 | 23 | 23 |
| Unchecked Low Level Calls | 3 | 4 | 7 |

```
-        uint256 _contractorProceeds = _conversionAmount - _maxReferralRewardETHWei - _moderatorFeeETHWei;
-        counters[1]++;
+        uint256 _contractorProceeds = _conversionAmount.sub(_maxReferralRewardETHWei.add(_moderatorFeeETHWei));
+        counters[1] = counters[1].add(1);
```

**Fig. 6** Arithmetic vulnerability fixed according to academic guidelines

lines. In contrast, some categories almost never matched guidelines: for instance, none of the 6 Bad Randomness instances or the 6 DOS instances matched, and for Time Manipulation, 0 out of 23 adhered to guidelines. Smaller categories like Front Running (1 out of 3), Short Address (1 out of 3), and Unchecked Low Level Calls (3 out of 7) also show limited guideline compliance.

The results confirm previous evidence in the literature. Arithmetic and reentrancy are not only the most frequent vulnerabilities but also among those with the highest adherence to literature guidelines, in line with findings by Durieux et al. (2020). Access control shows a similar trend, likely due to its overlap with the OWASP Top 10 and its general relevance to software security. By contrast, adherence is null for denial of service, bad randomness, and time manipulation, suggesting that current guidelines do not adequately cover these areas. Categories with fewer instances, such as front running and unchecked return values, exhibit only partial adherence.

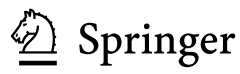

To add depth and inform readers about how the labeling phase worked, Fig. 6 depicts one instance of an arithmetic vulnerability fix that follows academic guidelines. As is visible, the use of *SafeMath* modifier according to prior gathered fixing guidelines disallows arithmetic vulnerability.

We do not discuss the percentage of unchecked return values for low-level calls and front-running vulnerabilities due to the limited number of fixing commits among those collected. For vulnerabilities where adherence to literature is 0, we subsequently highlight the innovative strategies that developers may employ as new security patterns or patches that are still emerging in response to fixes that have not yet received widespread attention in the research community.

At the end of this process, 221 commits contained fixing strategies already described in the literature, namely the 60.55% of the fixes in our sample. On the other hand, 143 commits introduced novel approaches. These unexplored fixes served as the foundation for the evaluation carried out in $\mathbf{RQ}_2$.

**Summary of Findings for $\mathbf{RQ}_1$**

The analysis reveals that developers significantly adhere to documented fixing strategies in the literature for vulnerabilities such as reentrancy, arithmetic, and access control, which are the most studied and well-understood due to their critical impact. In contrast, adherence is null for other types of vulnerabilities. This stark difference highlights the need to update existing guidelines to better mitigate less researched vulnerabilities, further motivating our study.

### 6.2 $\mathbf{RQ}_2$: New Fixes

To answer $\mathbf{RQ}_2$, we analyzed the 143 commits that did not align with any of the fixing strategies identified in the literature. These commits represent potential cases where developers applied alternative or novel solutions to known security issues in smart contracts.

From this analysis, we extracted 35 commits that introduced 27 distinct fixing strategies not previously described in academic work. These new strategies span multiple categories of the DASP taxonomy, with the following distribution: *Arithmetic Issues* (8), *Reentrancy* (5), *Denial of Service* (4), *Access Control* (2), *Front Running* (2), *Bad Randomness* (2), *Time Manipulation* (2), *Unchecked Return Values for Low-Level Calls* (1), and *Short Address Attack* (1).

This distribution highlights the vulnerability types where developers are most actively innovating in practice, often beyond what is covered in current academic guidelines.

As with the process followed in $\mathbf{RQ}_1$, the identification of these new strategies involved manual analysis by two authors per commit. When disagreements occurred, a third evaluator was involved in a consensus process. Overall, 15 conflicts were resolved, and inter-rater agreement prior to conflict resolution reached a Cohen's kappa value of 0.72, indicating substantial agreement.

In the remainder of this section, we present the newly identified fixing strategies, organized by vulnerability category. To avoid redundancy, similar or identical approaches are presented only once.

### 6.2.1 Access Control

This category includes 2 new fixes, which we have detailed below.

The commit in Fig. 7 addresses an access control issue. The function assumes that the caller (`msg.sender`) is always entitled to any remaining Ether in the contract. However, this assumption can be invalid in scenarios where multiple parties interact with the contract and the intended recipient of the refund is different from the caller, e.g., the refund should go to a predetermined address or the originator of a transaction, not the executor.

By requiring explicit addresses for refunds, the function avoids sending Ether to potentially unintended recipients. This change ensures that leftover Ether is sent to `callValueRefundAddress_`, an explicitly provided refund address, instead of `msg.sender`. This prevents unauthorized refunds and improves security by ensuring that the recipient is always defined by the caller.

The commit shown in Fig. 8 reported as a commit message "fix: add missing access control", in detail, changes made involved the addition of a custom modifier that we clarify below.

```
        function initiateArbitrumNativeBatch(
            address switchboardAddress_,
+           address callValueRefundAddress_,
+           address remoteRefundAddress_,
            ArbitrumNativeInitiatorRequest[]
                calldata arbitrumNativeInitiatorRequests_
        ) external payable {
  @@ -368,15 +370,17 @@ contract SocketBatcher is AccessControl {
                    arbitrumNativeInitiatorRequests_[index].packetId,
                    arbitrumNativeInitiatorRequests_[index].maxSubmissionCost,
                    arbitrumNativeInitiatorRequests_[index].maxGas,
-                   arbitrumNativeInitiatorRequests_[index].gasPriceBid
+                   arbitrumNativeInitiatorRequests_[index].gasPriceBid,
+                   callValueRefundAddress_,
+                   remoteRefundAddress_
                );
                unchecked {
                    ++index;
                }
            }

            if (address(this).balance > 0)
-               msg.sender.call{value: address(this).balance}("");
+               callValueRefundAddress_.call{value: address(this).balance}("");
        }
```

**Fig. 7** 1st access control new fix

```
-   function pullToken(address _token) external returns (uint256) {
+   function pullToken(address _token) external onlyWhitelistedExecutor returns (uint256) {
      uint256 prevBalance = totalAmount[_token];
      uint256 nextBalance = IERC20Upgradeable(_token).balanceOf(address(this));
```

**Fig. 8** 2nd access control new fix

```
modifier onlyWhitelistedExecutor()
    if (!serviceExecutors[msg.sender])
    revert IVaultStorage_NotWhiteListed();
    _;
}
```

With `serviceExecutors` defined as a mapping from address to bool.

```
mapping(address => bool) public serviceExecutors;
```

This commit adds essential access control to the `pullToken` function, ensuring that only authorized addresses can call it, checking if the transaction invoker is among the analyzed ones. This specific modifier grants access with fewer restrictions than the widely used `onlyOwner` and could be generally employed when dealing with functions with a permitted list of callers.

**Summary of Findings for Access Control fixes**

Authorize function requests using the address of the caller passed as an input parameter instead of using `msg.sender`. Use a mapping with addresses as keys and boolean as values to permit or not permit access to a given function. These strategies are other than those collected in the literature indicating the use of `onlyOwner` modifier.

### 6.2.2 Arithmetic

Within this category, we present 8 newly identified fixes.

The maximum penalty is the balance, by limiting the subtrahend to the max value of the minuend, underflow is actually fixed. The previous version allowed underflow `inactiveJuror.balance-inactiveJuror.atStake`and then underflow `inactiveJuror.balance`, which could have allowed an attacker to steal everything if he had managed to have `inactiveJuror.atStake> inactiveJuror.balance`. Figure 9 depicts the fix.

The function `.sub()` comes from *SafeMath* a common and known in the current literature way to securely deal with arithmetic operations without falling into overflows and underflows. This function returns an error in case of arithmetic issues. In the commit displayed in Fig. 10, the developer substituted the `.sub` with a custom function, namely, `subMax0`, which is codified as shown in Listing 1:

```
    function penalizeInactiveJuror(address _jurorAddress, uint _disputeID, uint[] _draws) public {
        Dispute storage dispute = disputes[_disputeID];
        Juror storage inactiveJuror = jurors[_jurorAddress];
        require(period > Period.Vote);
        require(dispute.lastSessionVote[_jurorAddress] != session); // Verify the juror hasn't voted.
        dispute.lastSessionVote[_jurorAddress] = session;
        require(validDraws(_jurorAddress, _disputeID, _draws));
        uint penality = _draws.length * minActivatedToken * 2 * alpha / ALPHA_DIVISOR;
-       penality = (penality < inactiveJuror.balance-inactiveJuror.atStake) ? penality : inactiveJuror.balance-inactiveJuror.atStake;
 what the juror can lose.
+       penality = (penality < inactiveJuror.balance) ? penality : inactiveJuror.balance; // Make sure the penality is not higher tha
        inactiveJuror.balance -= penality;
        jurors[msg.sender].balance += penality / 2; // Give half of the penalty to the caller.
        jurors[this].balance += penality / 2; // The other half to Kleros.

        msg.sender.transfer(_draws.length*dispute.arbitrationFeePerJuror);
    }
```

**Fig. 9** 1st arithmetic new fix

```
    function _getInterestValuePerLP(uint256 expiry, address account)
        internal
        override
        returns (uint256 interestValuePerLP)
    {
        ExpiryData storage exd = expiryData[expiry];
        if (userLastNormalizedIncome[expiry][account] == 0) {
            interestValuePerLP = 0;
        } else {
-           interestValuePerLP = exd.paramL.sub(
+           interestValuePerLP = exd.paramL.subMax0(
                exd.lastParamL[account].mul(globalLastNormalizedIncome[expiry]).div(
                    userLastNormalizedIncome[expiry][account]
                )
            );
        }
```

**Fig. 10** 2nd arithmetic new fix

**Listing 1** subMax0 function

```
1 function subMax0(uint256 a, uint256 b) internal pure
2     returns (uint256) {
3     return a >= b ? a - b : 0;
4 }
```

In Solidity, `uints` are unsigned integers, thus, variables of this type cannot represent negative values. The proposed fix assigns 0 if the value becomes negative, without returning an error message. This fix is particularly suitable when dealing with units that, according to the business logic can have 0 as a minimum value. The commit message reports "*Fix bug of possible overflow subtraction in Aave LiqMining and Market*". Where `AaveLiquidityMining` and `Market` are two contracts that were both involved in the same changes. These contracts extend the same base contract, `PendleLiquidityMiningBase`, and override the function `_getInterestValuePerLP`.

The function `_getInterestValuePerLP` is not called directly by external users, but it is internally used in the interest settlement mechanism. Specifically, it is first called through a `for` loop inside the function `claimLpInterests()`, which in turn invokes `_settleLpInterests` as illustrated in Listing 2.

The full call chain is the following:

$$\text{claimLpInterests} \longrightarrow \text{_settleLpInterests} \longrightarrow \text{_getInterestValuePerLP}$$

**Listing 2** claimLpInterests and _settleLpInterests functions

```
function claimLpInterests() external override
    nonReentrant returns (uint256 interests) {
    for (uint256 i = 0; i < userExpiries[msg.sender].
        expiries.length; i++) {
        interests = interests.add(
            _settleLpInterests(userExpiries[msg.sender].
                expiries[i], msg.sender)
        );
    }
}

function _settleLpInterests(uint256 expiry, address
    account)
    internal
    returns (uint256 dueInterests)
{
    ExpiryData storage exd = expiryData[expiry];

    if (account == address(exd.lpHolder)) return 0;

    _updateParamL(expiry);

    uint256 interestValuePerLP = _getInterestValuePerLP(
        expiry, account);
    ...
}
```

In this scenario, if one uses the standard `sub()` and any subtraction results in a negative value, all operations performed within the loop will be reverted. This behavior is avoided by returning 0 as a result.

Relying only on SafeMath to handle arithmetic vulnerabilities may be a limit. In some cases, if the logic is not correct, the contracts will return errors without functioning. On the other hand, using the arithmetic default check of Solidity 0.8+ will cause reverts. In the commit diff shown in Fig. 11, if the contract already holds some ETH before the swap, `address(this).balance` includes this existing balance. Swap serves to obtain the exact amount of LUSD needed to repay the debt by swapping collateral or other tokens. If `collateralReturned`(which is `address(this).balance`) is greater than `collToWithdraw` due to the existing balance, the subtraction:

$$\text{collateralSold} = \text{collToWithdraw} - \text{collateralReturned}$$

results in an underflow. This causes `collateralSold` to wrap around to a very large number, leading to incorrect logic flow.

```
+        // Saving a balance to a local variable to later get a `collateralReturned` value unaffected by a previously
+        // held balance --> This is important because if we set `collateralReturned` to `address(this).balance`
+        // the value might be larger than `collToWithdraw` which could cause underflow when computing `collateralSold`
+        uint256 ethBalanceBeforeSwap = address(this).balance;
+
         (bool success, ) = LUSD_USDC_POOL.call(
             abi.encodeWithSignature(
                 "swap(address,bool,int256,uint160,bytes)",
                 address(this), // recipient
                 false, // zeroForOne
                 -int256(lusdToBuy),
                 SQRT_PRICE_LIMIT_X96,
                 abi.encode(SwapCallbackData({debtToRepay: debtToRepay, collToWithdraw: collToWithdraw}))
             )
         );

         if (success) {
             // Note: Debt repayment took place in the `uniswapV3SwapCallback(...)` function
-            collateralReturned = address(this).balance;
+            collateralReturned = address(this).balance - ethBalanceBeforeSwap;

             {
                 // Check that at most `maxCost` of ETH collateral was sold for `debtToRepay` worth of LUSD
                 uint256 maxCost = (lusdToBuy * _maxPrice) / PRECISION;
                 uint256 collateralSold = collToWithdraw - collateralReturned;
                 if (collateralSold > maxCost) revert MaxCostExceeded();
             }
```

**Fig. 11** 3rd arithmetic new fix

In the updated function, the ETH balance is stored before the swap, and `collateralReturned` is calculated based on the difference with the balance before the swap:

$$\text{ethBalanceBeforeSwap} = \text{address(this).balance}$$

$$\text{collateralReturned} = \text{address(this).balance} - \text{ethBalanceBeforeSwap}$$

This prevents incorrect execution paths, such as unintended reverts or security breaches due to manipulated `collateralSold` values. It ensures that the calculations accurately reflect only the ETH received from the swap, enhancing the security and reliability of the contract.

The changes made in the commit in Fig. 12 address the overflow bug by introducing boundary checks to ensure that `positionInArray` does not exceed the length of

```
         // Delete entries of helper directories.
         uint256[] memory arrayMem = topics2ClaimIds[claim.topic];
         topics2ClaimIds[claim.topic].length = 0;
         topics2ClaimIds[claim.topic].length = arrayMem.length - 1;
         uint256[] memory arrayStor = topics2ClaimIds[claim.topic];
         uint32 positionInArray = 0;
-        while(_claimId != arrayMem[positionInArray]) {
+        while(positionInArray < arrayMem.length && _claimId != arrayMem[positionInArray]) {
             positionInArray++;
         }

+        // Make sure that the element has actually been found.
+        require(positionInArray < arrayMem.length);
+
```

**Fig. 12** 4th arithmetic new fix

```
          // calc amt{0,1} for the change of liquidity
-         uint128 absL = uint128(uint96(liquidityDeltaD8 >= 0 ? liquidityDeltaD8 : -liquidityDeltaD8) << 8);
+         uint128 absL = uint128(uint96(liquidityDeltaD8 >= 0 ? liquidityDeltaD8 : -liquidityDeltaD8)) << 8;
```

**Fig. 13** 5th arithmetic new fix

```
          // 0.9999e18 accounts for rounding errors in the Uniswap V3 protocol.
-         uint128 amount0Min = amount0 == 0 ? 0 : (amount0 * 0.9999e18) / 1e18;
-         uint128 amount1Min = amount1 == 0 ? 0 : (amount1 * 0.9999e18) / 1e18;
+         uint128 amount0Min = amount0 == 0 ? 0 : (amount0 * 0.9999e4) / 1e4;
+         uint128 amount1Min = amount1 == 0 ? 0 : (amount1 * 0.9999e4) / 1e4;
```

**Fig. 14** 6th arithmetic new fix

`arrayMem`. Without this check, if `_claimId` is not present in `arrayMem`, `positionInArray` would continue incrementing indefinitely, potentially causing an array out-of-bounds access or an overflow of `positionInArray`. By adding the condition `positionInArray` < `arrayMem.length`, the loop exits when `_claimId` is not in the array, thus preventing `positionInArray` from surpassing the array's bounds.

The commit changes represented in Fig. 13 fix an overflow vulnerability. If `liquidityDeltaD8` is positive or zero, use its value. If it is negative, use its opposite (the absolute value). If the result exceeds $2^{96} - 1$ (the maximum value for `uint96`), an overflow may occur during the cast to `uint96`. By casting to `uint96` before shifting, we ensure that `liquidityDeltaD8` fits within 96 bits.

Shifting a `uint96` value left by 8 bits results in a value that fits within 104 bits, which is safely accommodated by the final cast to `uint128`.

In Solidity 1e18 means $1 \times 10^{18}$. In code before the commit shown in Fig. 14, is calculated `amount0Min` and `amount1Min` by multiplying `amount0` and `amount1` by $0.9999 * 10^{18}$ (written as `0.9999e18`) and then dividing by $1 * 10^{18}$ (written as `1e18`). This approach was intended to compute 99.99% of `amount0` and `amount1`. However, when dealing with large numbers of type `uint128`, multiplying them by `0.9999e18` could cause an overflow because the intermediate result becomes too large to fit within a `uint128` variable. To fix this issue, the updated code changes the scaling factors from `0.9999e18` and `1e18` to `0.9999e4` and `1e4`. Now, they multiply by $0.9999 * 10^4$ (i.e., `0.9999e4`) and divide by $1 * 10^4$ (i.e., `1e4`). This adjustment still computes 99.99% of `amount0` and `amount1`, but using much smaller numbers.

By scaling down the factors, the intermediate multiplication results remain within the safe range of a `uint128`, mitigating overflow. This change preserves the original intent of calculating 99.99% of the amounts while ensuring the calculations are safe for large values.

The issue highlighted in Fig. 15 is that the line `store. decrementBufferBalance(amount);`is executed before checking if `amount >bufferBalance`. If `amount` is greater than `bufferBalance`, subtracting `amount` from `bufferBalance` could cause an arithmetic underflow. The updated code first checks if `amount` is greater than `bufferBalance`. Only if `amount` is less than or equal to `bufferBalance` will the code proceed to `store .decrementBufferBalance(amount).`

```
-        store.decrementBufferBalance(amount);
-
         if (amount > bufferBalance) {
             uint256 diffToPayFromPool = amount - bufferBalance;
             uint256 poolBalance = store.poolBalance();
             require(diffToPayFromPool < poolBalance, "!pool-balance");
             store.decrementPoolBalance(diffToPayFromPool);
+        } else {
+            store.decrementBufferBalance(amount);
         }
```

**Fig. 15** 7[th] arithmetic new fix

```
     for (uint256 i; i < solverOps.length; ++i) {
         bidAmountFound = _getBidAmount(dConfig, userOp, solverOps[i], returnData, key);

-        if (bidAmountFound == 0) {
+        if (bidAmountFound == 0 || bidAmountFound > type(uint240).max) {
             // Zero bids are ignored: increment zeroBidCount offset
+            // Bids that would cause an overflow are also ignored
             unchecked {
                 ++zeroBidCount;
             }
         } else {
-            if (bidAmountFound > type(uint240).max) revert BidTooHigh(i, bidAmountFound);
-
```

**Fig. 16** 8[th] arithmetic new fix

By performing the check first, the code ensures that the `decrementBufferBalance` operation is only called when there is enough balance in the buffer. This is a different fix compared to using `SafeMath`, as it relies on explicit conditional checks and different order of operations to prevent underflows rather than using a library to handle arithmetic safety.

The snippet depicted in Fig. 16 is extracted from a contract with Solidity 0.8.22; in Solidity 0.8.0+, overflow and underflow checks are enabled by default, causing a revert. In the initial code, if `bidAmountFound` was greater than `type(uint240).max`, the contract would revert with an error `BidTooHigh`. This implies that the entire operation would fail if a bid amount is too high, influencing all the other operations executed in the for cycle.

In the fixed code, instead of reverting, bids that would cause an overflow (`bidAmountFound >type(uint240).max`) are now ignored. The logic increments `zeroBidCount` to treat these bids as zero bids, allowing the operation to continue smoothly.

This fix suggests that protection mechanisms based on SafeMath are being used less frequently, as Solidity 0.8.0+ includes built-in overflow and underflow checks by default. However, previous research has shown that the import of SafeMath was historically the most frequently used OpenZeppelin import (Khan et al. 2022). On the other hand, Wang et al. showed that although many new features are introduced and deprecated ones removed, not all changes necessarily work in favor of the developers (Wang et al. 2021).

```
                getEmissionDataForUser[currentIdOwner].emissionMultiple += uint64(newCurrentIdMultiple);

                // Update the random seed to choose a new distance for the next iteration.
-               currentRandomSeed = uint256(keccak256(abi.encodePacked(currentRandomSeed)));
+               // It is critical that we cast to uint64 here, as otherwise the random seed
+               // set after calling revealGobblers(1) thrice would differ from the seed set
+               // after calling revealGobblers(3) a single time. This would enable an attacker
+               // to choose from a number of different seeds and use whichever is most favorable.
+               currentRandomSeed = uint64(uint256(keccak256(abi.encodePacked(currentRandomSeed))));
            }
```

**Fig. 17** 1st bad randomness new fix

This raises interesting questions about how developers are adapting to these changes and whether they are fully leveraging Solidity's built-in protections. Future work should further investigate these aspects, examining whether SafeMath is still being used in certain contexts.

> **Summary of Findings for Arithmetic fixes**
>
> To prevent underflow, limit the subtrahend to the max value of the minuend. Report 0 instead of a require error message if 0 is a valid result for a given operation, and the requirement error propagation would break a loop. Before performing swaps save a local balance to perform further operations. Check counter overflow when looping over an array. Properly shift buffers when using typecasting at the end of the chain of arithmetic operations. Properly scale uint variable to be sure that these fit in the buffer. Gathered fixing procedures that suggest the use of SafeMath or require statements to check overflows and underflows.

### 6.2.3 Bad Randomness

We identified 2 previously undocumented fixes in this category, which we discuss in the following.

The code change shown in Fig. 17 fixes a randomness flaw that could allow an attacker to generate multiple random seeds and select the best outcome. By casting the seed to `uint64`, the random seed remains consistent no matter how the function is called, reducing the potential for manipulation and making the randomness harder to exploit. The developer's comment summarizes the justification for this fix.

In the code shown in Fig. 18, `msg.sender` is used as part of the input to generate randomness, before the update made by the commits. `msg.sender` is the address of the caller

```
                    tokenId < metadatas[i].endIndex
                ) {
                    randomness = uint256(
-                       keccak256(
-                           abi.encode(metadatas[i].entropy, msg.sender, tokenId)
-                       )
+                       keccak256(abi.encode(metadatas[i].entropy, tokenId))
                    );
                    metadataCleared = true;
                }
```

**Fig. 18** 2nd bad randomness new fix

of the contract. An attacker knows that their own address (`msg.sender`) is included in the randomness calculation, and they can potentially influence the result. For example, they could call the contract repeatedly with different addresses (or from different wallets) until they get a desired outcome, thus manipulating the randomness. The randomness generation becomes less dependent on variables that can be controlled or influenced by an external party, mitigating bad randomness.

> **Summary of Findings for Bad Randomness fixes**
>
> Disallow malicious users to choose among different seeds not to enable them to pick the most favorable. Remove all the randomness sources that may be controlled or known by attackers.

### 6.2.4 Denial of Service

In this category, we identified 4 new fixes. The fix highlighted in Fig. 19 addresses Denial of Service vulnerability pattern that arises from repeated, unnecessary actions on the same state. The introduced check returns an error if a claim has already been requested to prevent DoS vulnerabilities caused by redundant operations.

In Fig. 20, the added line under the comment sets an upper limit on the number of reward tokens processed in order to prevent a DOS attack. Without the added line, iterating too many times could consume excessive gas and make the transaction fail.

The fixing strategy underscored in Fig. 21 disallows a DOS attack. The function `onRepay` checks how much of the daily limit the user has used, ensuring fair access to borrowing for all users and preventing one user from denying service through repeated borrow-repay cycles.

```
+           challengeDefended = false;
            err = this.bondDeposit(superblockHash, msg.sender, minProposalDeposit);
            require(err == ERR_SUPERBLOCK_OK);

 @@ -227,6 +228,10 @@ contract SyscoinClaimManager is Initializable, SyscoinDepositsM
                emit ErrorClaim(superblockHash, ERR_SUPERBLOCK_BAD_CLAIM);
                return (ERR_SUPERBLOCK_BAD_CLAIM, superblockHash);
            }
+           if(challengeDefended == true){
+               emit ErrorClaim(superblockHash, ERR_SUPERBLOCK_CLAIM_ALREADY_DEFENDED);
+               return (ERR_SUPERBLOCK_CLAIM_ALREADY_DEFENDED, superblockHash);
+           }
            if (claim.decided || claim.invalid) {
                emit ErrorClaim(superblockHash, ERR_SUPERBLOCK_CLAIM_DECIDED);
                return (ERR_SUPERBLOCK_CLAIM_DECIDED, superblockHash);
```

**Fig. 19** $1^{st}$ DoS new fix

The patch to DOS vulnerability reported in Fig. 22 sets a minimum value for each deposit of 1 ether, which is a valuable amount. Establishing a high minimum deposit for each transaction prevents attackers from successfully denying service to a specific contract.

**Summary of Findings for Denial of Service fixes**

To address DoS vulnerabilities caused by redundant operations, track the state of each operation or entity and validate the state before allowing subsequent actions. Set an upper limit while looping. Set a temporal limit needed to recall a given function. Set a minimum deposit to discourage attackers from repeatedly calling an SC function. The collected approaches are diverse from barely avoiding using `transfer()` in loops.

### 6.2.5 Front-Running

This category includes 2 new fixes, that we detail below.

In the previous version of the contract, the salt is derived using: `bytes32 salt = keccak256(abi.encode(owner))`.

Here, the salt is only dependent on the `owner` address. This predictability allows a malicious actor to see the transaction and, if advantageous, front-run the transaction by submit-

```
    function _checkpoint(address _account, bool claim) internal {
        uint256 supply = managed_assets;
        uint256 depositedBalance;
        depositedBalance = aggregatedAssetsOf(_account);

        IRewardStaking(convexPool).getReward(address(this), true);

        uint256 rewardCount = rewards.length;
+       // Assuming that the reward distribution takes am avg of 230k gas per reward token we are setting an
 upper limit of 40 to prevent DOS attack
+       rewardCount = rewardCount >= 40 ? 40 : rewardCount;
        for (uint256 i; i < rewardCount; ++i) {
            _calcRewardIntegral(i, _account, depositedBalance, supply, claim);
        }
    }
```

**Fig. 20** 2nd DoS new fix

```
+    function onRepay(uint amount) public {
+        uint day = block.timestamp / 1 days;
+        if(dailyBorrows[msg.sender][day] < amount) {
+            dailyBorrows[msg.sender][day] = 0;
+        } else {
+            dailyBorrows[msg.sender][day] -= amount;
+        }
+    }
  }
```

**Fig. 21** 3rd DoS new fix

```
@@ -11,6 +11,9 @@ uint256 constant VALIDATORS_MAX_AMOUNT = 400;
  /// @dev Collateral size of 1 validator
  uint256 constant COLLATERAL = 32 ether;

+ /// @dev Minimal 1 time deposit
+ uint256 constant MIN_DEPOSIT = 1 ether;
+
  /// @dev Lockup time to allow P2P to make ETH2 deposits
  /// @dev If there is leftover ETH after this time, it can be refunded
  uint40 constant TIMEOUT = 1 days;
```

**Fig. 22** 4th DoS new fix

ting a similar one with the same predictable salt, but with a higher gas price, ensuring their transaction is processed first. The fix used in Fig. 23 introduces `tx.origin` into the salt computation. `tx.origin` is the original sender of the transaction, even if multiple contract calls are involved. The salt becomes tied to the original transaction initiator, even if an attacker sees the transaction, they cannot simply replicate or predict the salt unless they are the original sender.

The `onlyGovernance` modifier restricts certain functions so that only the governance entity (e.g., a *multisig* wallet, or *DAO*) can call them. Here is how it typically functions:

```
-     function deployFor(address owner) public override returns (address payable proxy) {
-         // Deploy the proxy contract with CREATE2.
+     function deployFor(address owner, bytes32 salt) public override returns (address payable proxy) {
+         // Load the proxy bytecode.
          bytes memory bytecode = type(PRBProxy).creationCode;
-         bytes32 salt = keccak256(abi.encode(owner));
+
+         // Prevent front running the salt by hashing the concatenation of msg.sender and the user-provided salt.
+         salt = keccak256(abi.encode(tx.origin, salt));
+
+         // Deploy the proxy with CREATE2.
          assembly {
              let endowment := 0
              let bytecodeStart := add(bytecode, 0x20)
              let bytecodeLength := mload(bytecode)
              proxy := create2(endowment, bytecodeStart, bytecodeLength, salt)
          }

          // Transfer the ownership from this factory contract to the specified owner.
          IPRBProxy(proxy)._transferOwnership(owner);

          // Mark the proxy as deployed in the mapping.
          isProxy[proxy] = true;

          // Log the proxy via en event.
          emit DeployProxy(msg.sender, owner, address(proxy));
      }
  }
```

**Fig. 23** 1st front-running new fix

```
1
2 modifier onlyGovernance() {
3     require(msg.sender == governanceManager, "
        Not_governance");
4     _;
5 }
```

**Listing 3** onlyGovernance Modifier.

The change in the commit shown in Fig. 24 is the order of the modifiers `initializer` and `onlyGovernance`. Specifically, the order was changed from `onlyGovernance initializer` to `initializer onlyGovernance`. This change is important because of how Solidity processes modifiers, which are processed in order.

Proxied contracts do not make use of a constructor, it is indeed common to move constructor logic to an external initializer function. It then becomes necessary to protect this initializer function so it can only be called once to prevent reinitializations.

The `initializer` modifier in this contract comes from *OpenZeppelin Initialiable*. It ensures that the `initialize` function can only be called once. By placing `initializer` before `onlyGovernance`, the contract ensures that the `initializer` modifier's logic is executed first. This prevents any other action from being taken before the `initializer` check is enforced, then it checks the permission to call the function.

**Summary of Findings for Front Running fixes**

Prevent the manipulation of the transaction order by relating the transaction to specific users, using their provided and self-known data. To avoid leaving the proxy in an uninitialized state, the initializer function should be called as early as possible, making the initialization done just one time by disallowing attackers to *front-run* logic in other modifiers before initialization. Such strategies vary from requiring the current allowance to match the expected value or zero that we found in the literature.

### 6.2.6 Reentrancy

Within this category, we present 5 newly identified fixes. Evidence in the literature treats deeply reentrancy when dealing with token transfer (Chen et al. 2020; Zhou et al. 2023a; Chen et al. 2023a). This vulnerability can also occur with other kinds of state manipulations, as underscored in Fig. 25, which are less considered. The Checks-Effects-Interactions pattern results even in this case a valid mitigation. By deleting or updating the state variables

```
    function initialize(
        address newVePendle,
        address newGaugeController,
        bytes memory _marketCreationCode
-   ) external onlyGovernance initializer {
+   ) external initializer onlyGovernance {
        require(newVePendle != address(0) && newGaugeController != address(0), "zero address");
        vePendle = newVePendle;
        gaugeController = newGaugeController;
        marketCreationCodePointer = SSTORE2Deployer.setCreationCode(_marketCreationCode);
    }
```

**Fig. 24** 2[nd] front-running new fix

```
    function finalizeRecovery(BaseWallet _wallet) external onlyWhenRecovery(_wallet) {
        RecoveryConfig storage config = recoveryConfigs[address(_wallet)];
        require(uint64(now) > config.executeAfter, "RM: the recovery period is not over yet");
-       _wallet.setOwner(config.recovery);
-       emit RecoveryFinalized(address(_wallet), config.recovery);
-       guardianStorage.setLock(_wallet, 0);
+       address recoveryOwner = config.recovery;
        delete recoveryConfigs[address(_wallet)];
+
+       _wallet.setOwner(recoveryOwner);
+       guardianStorage.setLock(_wallet, 0);
+
+       emit RecoveryFinalized(address(_wallet), config.recovery);
    }
```

**Fig. 25** $1^{st}$ reentrancy new fix

before making any external calls, the contract ensures that even if a reentrancy attack is attempted, the critical state has already been modified, and the attacker cannot exploit the previous state.

Operating with ERC777 from version 3.3.0 or earlier, and defining a custom `_beforeTokenTransfer`function that writes to a storage variable, may be vulnerable to a reentrancy attack. One characteristic of ERC777 is that it permits reentrancy through the send-and-receive hooks. Therefore, the token must be programmed carefully to prevent a reentrancy attack. Specifically, the contract should be consistent whenever an external call is made to an untrusted address. When burning tokens, the function `_beforeTokenTransfer` is called before the transfer hook is activated for the sender. While the token balances are adjusted after this function is executed, there is a moment during the call to the sender where reentrancy could occur. At this point, the state managed by `_beforeTokenTransfer` may not reflect the actual token balances or the total supply.

The fix reported in Fig. 26 addresses the described issue by calling the custom `_beforeTokenTransfer` after changing the state of the contract.

The fixing strategy in Fig. 27 introduces a new variable `beforeNFTBalance` to capture the NFT balance of the recipient before transferring tokens. It also added a *require* statement to ensure that the NFT balance remains unchanged, protecting against reentrancy, and ensuring that a reentrant call cannot manipulate the NFT balance and perform an attack.

Functions like `.transfer()` and `.send()` have often been proposed as valid reentrancy fixes (Zhou et al. 2023a). The behavior underlying the mitigation relies on limiting the amount of gas forwarded to the called contract. Specifically, both `.transfer()` and `.send()` forward only 2300 gas to the recipient. This amount of gas is insufficient to execute complex operations, such as reentering the vulnerable contract and making further external calls. This guidance made sense under the assumption that gas costs would not change, but that assumption turned out to be incorrect. Indeed, each opcode supported by the EVM has an associated gas cost that could change, so SCs should not depend on any particular gas costs, as they do with `.send()` and `.transfer()`. Therefore, if the gas cost changes, these changes could enable reentrancy.

As Fig. 28 shows, it is recommended to use `.call()` when there are no state changes involved, or when the function has a lock, a *nonReentrant* modifier, or follows the Checks-Effects-Interactions Pattern. This avoids reentrancy considering long-term effects.

```
    function _burn(
        address from,
        uint256 amount,
        bytes memory data,
        bytes memory operatorData
    )
        internal
        virtual
    {
        require(from != address(0), "ERC777: burn from the zero address");

        address operator = _msgSender();

-       _beforeTokenTransfer(operator, from, address(0), amount);
-
        _callTokensToSend(operator, from, address(0), amount, data, operatorData);

+       _beforeTokenTransfer(operator, from, address(0), amount);
+
        // Update state variables
        _balances[from] = _balances[from].sub(amount, "ERC777: burn amount exceeds balance");
        _totalSupply = _totalSupply.sub(amount);

        emit Burned(operator, from, amount, data, operatorData);
        emit Transfer(from, address(0), amount);
    }
```

**Fig. 26** 2nd reentrancy new fix

```
            // Call router to transfer tokens from user
            uint256 beforeBalance = _token.balanceOf(_assetRecipient);
 -
 +          uint256 beforeNFTBalance = nft().balanceOf(_assetRecipient);
            router.pairTransferERC20From(
                _token,
                routerCaller,
@@ -71,6 +71,10 @@ abstract contract LSSVMPairERC20 is LSSVMPair {
                    inputAmount,
                "ERC20 not transferred in"
            );
 +          require(
 +              beforeNFTBalance == nft().balanceOf(_assetRecipient),
 +              "Reentrant call from router"
 +          );
        } else {
            // Transfer tokens directly
            _token.safeTransferFrom(msg.sender, _assetRecipient, inputAmount);
```

**Fig. 27** 3rd reentrancy new fix

The removed line in Fig. 29 checks whether the contract is in an "executing" state using a boolean flag `isExecuting`. This flag is meant to ensure that the `executeWithdrawOrder` function can only be executed when the contract is in a specific state.

Ensuring that the function can only be executed by the contract itself, prevents external attackers from directly calling this function in a way that could manipulate the contract's

```
    /// @inheritdoc IKrStaking
    function rescueNative() external payable onlyRole(OPERATOR_ROLE) {
-       payable(msg.sender).transfer(address(this).balance);
+       (bool success, ) = msg.sender.call{value: address(this).balance}("");
+       require(success, "Transfer failed.");
    }
```

**Fig. 28** 4th reentrancy new fix

state maliciously. This pattern makes it impossible for an attacker to execute the function through a fallback or reentrant call from an external contract.

> **Summary of Findings for Reentrancy fixes**
>
> Delete or update state variables before external interactions. Check if the previous balance is unchanged before updating the state of the contract. When using ERC777 from version 3.3.0 or earlier, use custom *beforeTokenTransfer* after state changes. Use *call()* instead `transfer()` it *send()* if the function does not update the state, follow the Checks-Effects-Interactions pattern or use locking mechanisms. Ensure that reentrancy-prone functions can only be invoked by the contract itself. hese approaches diverge from those found in the literature reviewed by differing from using OpenZeppelin modifiers, using standard patterns, and avoiding the use of call().

### 6.2.7 Short Address

We identified 1 new fix in this category. The commit displayed in Fig. 30 patches a short address vulnerability. The `transfer` function in the original code does not check the size of the payload in `msg.data`, making it vulnerable to the short-address attack. The `onlyPayloadSize` modifier checks the length of the `msg.data` and ensures it is the expected size for the `transfer` function.

`msg.data.length` is the length of the input data for the transaction. The expected size for the `transfer` function parameters is `2 * 32` bytes (since both `address` and `uint` are 32 bytes each), plus an extra 4 bytes for the function selector.

`assert(msg.data.length == size + 4);` ensures that the transaction data has the correct length. If the length is incorrect, the transaction will be reverted, preventing a short address attack. By validating the size of `msg.data`, the `onlyPayloadSize` modifier ensures that the parameters passed to the `transfer` function are of the expected

```
  function executeWithdrawOrder(WithdrawOrder memory _order) external {
    // if not in executing state, then revert
-   if (!isExecuting) revert ICrossMarginHandler_NotExecutionState();
+   if (msg.sender != address(this)) revert ICrossMarginHandler_NotExecutionState();

    // Call service to withdraw collateral
    if (_order.shouldUnwrap) {
```

**Fig. 29** 5th reentrancy new fix

```
-   function transfer(address _to, uint _value) {
+ /*
+  * Fix for the ERC20 short address attack
+  */
+   modifier onlyPayloadSize(uint size) {
+      assert(msg.data.length == size + 4);
+      _;
+   }
+
+   function transfer(address _to, uint _value) onlyPayloadSize(2 * 32) {
        balances[msg.sender] = safeSub(balances[msg.sender], _value);
        balances[_to] = safeAdd(balances[_to], _value);
        Transfer(msg.sender, _to, _value);
```

**Fig. 30** 1st short address new fix

length. This prevents malicious actors from providing a shortened address that could lead to incorrect value calculations or balance updates.

> **Summary of Findings for Short Address fixes**
>
> To mitigate short address attacks, developers should validate the payload size of transaction data (`msg.data.length`) using a modifier such as `onlyPayloadSize`. This ensures that input parameters match the expected length, preventing malicious actors from exploiting truncated addresses to manipulate value transfers or state updates.

### 6.2.8 Time Manipulation

Within this category, we present 2 new fixes. Figure 31 displays a time manipulation fixing approach, which is reached by avoid to rely on `now` and using as a timestamp a value passed as an input of the function. In Solidity, `now` is an alias of `block.timestamp` which could be manipulated by the miners, relying on a timestamp received in input or from a trusted oracle the issue is solved. Notice that `_startDate` stands out as an input parameter instead of a variable created and initialized in the function, and `startDate` is a state variable[4].

The Yellow Paper[5] does not have any answer to "how much can it be off before it is rejected by other nodes". If `block.timestamp` is used, the only guarantee (equation 43) is that `block.timestamp` is greater than that of its parent. Ethereum clients like Geth and Parity reject blocks if their timestamps are more than 15 seconds in the future, therefore, this is the temporal window that may permit the validation of manipulated blocks. This implies that one can safely use `block.timestamp` if the time-dependent logic can tolerate a potential variation of up to 15 seconds.

[4] https://github.com/gnosis/pm-contracts/commit/81b40df2fe17dbcf1e4c65d7e2f946fed23cb351

[5] https://ethereum.github.io/yellowpaper/paper.pdf

```
    function StandardMarketWithPriceLogger(address _creator, Event _eventContract, MarketMaker _marketMaker, uint24 _fee, uint _startDate)
        public
        StandardMarket(_creator, _eventContract, _marketMaker, _fee)
    {
        if (_startDate == 0)
            startDate = now;
        else {
            // The earliest start date is the market creation date
-           require(startDate >= now);
+           require(_startDate >= now);
            startDate = _startDate;
        }
    }
```

**Fig. 31** 1st time manipulation new fix

The fix shown in Fig. 32 involves the removal of `require(nextMint [_localFarmAddress] < block.timestamp);` which is manipulable by the miners. The reported fix introduces a more sophisticated time-checking mechanism that ensures that minting rewards can only occur if a sufficient duration `(rewardDuration)` has passed since the last minting event. Thus, it prevents unauthorized reward minting for timeframes that could be manipulated, since Ethereum miners can only slightly influence the value of `block.timestamp`.

> **Summary of Findings for Time Manipulation fixes**
>
> Since *now* and `block` properties are prone to be manipulated by the miners, to mitigate time manipulation attacks, it is indicated to rely on timestamps passed as input parameters in the function that uses them. To use `block.timestamp` when a 15-second variance in time is acceptable for your application.

### 6.2.9 Unchecked Return Values for Low Level Call

This category includes one new fix that we detail below. The function `transfer(address _to, uint256 _value)` is included in the IERC20 interface provided by OpenZeppelin. According to the docs, it returns a boolean value indicating whether the operation succeeded. In the context of ERC20 tokens, the `transfer` and `transferFrom` functions are essential for transferring tokens between accounts. These functions are designed to return a boolean value indicating whether the operation was successful. However, many smart contracts interacting with ERC20 tokens often assume that `transfer` and `transferFrom` will always succeed and do not check the returned boolean value. This assumption can create a false sense of security, as a transfer may fail without the contract recognizing it. Such oversight can lead to incorrect token balances and other contract state inconsistencies.

SafeTransfer used as a fix comes from SafeERC20.sol of OpenZeppelin, which provides a wrapper around the standard ERC20 functions and handles the returns values. Such a fix is depicted in Fig. 33.

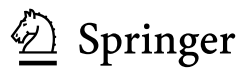

```
-
-        uint256 amount = tokensPerYear * _period / 365 days; // for all farms
-        amount = amount * localFarms[localFarmId[_localFarmAddress]].multiplier / totalMultipliers; // amour
-
-        nextMint[_localFarmAddress] = nextMint[_localFarmAddress] + _period;
-
-        rewardsToken.mint(_localFarmAddress, amount);
-        ILocalFarm(_localFarmAddress).notifyRewardAmount(amount);
+        require (_period > 0, "Cannot claim reward for a timeframe of 0 seconds");
+        //require (nextMint[_localFarmAddress] < block.timestamp); // Can not place a "requirement" on auto-
+
+        if(localFarms[localFarmId[_localFarmAddress]].lastMintTimestamp + rewardDuration < block.timestamp)
+        {
+            // We should check if sufficient time since the last minting session passed for this Local Farm.
+
+            if(_period < block.timestamp - localFarms[localFarmId[_localFarmAddress]].lastMintTimestamp)
+            {
+                // Claiming for less-than-maximum period.
+                // This can be necessary if the contract stayed without claiming for too long
+                // and the accumulated reward can not be minted in one transaction.
+
+                // In this case this function `mintFarmingReward` can be manually claimed by a user multiple
+                // in order to mint reward part-by-part.
+
+                // Last Mint Timestamp of the local farm must be updated to match the time preriod
+                // that user already claimed reward for.
+
+                localFarms[localFarmId[_localFarmAddress]].lastMintTimestamp += _period;
+            }
+            else
+            {
+                // Otherwise the reward is distributed for the total reward period duration,
+                // it is important to note that user can not cause the contract to print reward in the futu
+
+                _period = block.timestamp - localFarms[localFarmId[_localFarmAddress]].lastMintTimestamp;
+                localFarms[localFarmId[_localFarmAddress]].lastMintTimestamp = block.timestamp;
+            }
+
+            // Reward is then calculated based on the _period
+            // it can be either "specified period" or "now - Local Farm's previous minting timestamp".
+            uint256 amount = tokensPerYear * _period / 365 days; // for all farms
+            amount = amount * localFarms[localFarmId[_localFarmAddress]].multiplier / totalMultipliers; // a
+
+            // Local farm is then notified about the reward minting session.
+            rewardsToken.mint(_localFarmAddress, amount);
+            ILocalFarm(_localFarmAddress).notifyRewardAmount(amount);
+        }
```

**Fig. 32** 2nd time manipulation new fix

This fixing procedure diverges from barely checking with an if or a require statement the return value of the low-level call, by using an external library function.

> **Summary of Findings for Unchecked Return Values fixes**
>
> To mitigate risks from unchecked return values in ERC20 transfers, developers should avoid assuming that `transfer` and `transferFrom` always succeed. Instead of relying solely on conditional checks, the use of OpenZeppelin's `SafeERC20` library is recommended, as it wraps token operations and enforces proper handling of return values, ensuring safer and more reliable transactions.

```
        ITWAPOracle(manager.twapOracleAddress()).update();
        uint256 toTransfer =
            IERC20(manager.curve3PoolTokenAddress()).balanceOf(address(this));
-       IERC20(manager.curve3PoolTokenAddress()).transfer(
+       IERC20(manager.curve3PoolTokenAddress()).safeTransfer(
            manager.treasuryAddress(), toTransfer
        );
        emit PriceReset(
```

**Fig. 33** 1st unchecked return values for low level call new fix

### 6.2.10 Summary of Novel Fixes

To improve clarity and readability, we provide in Table 6 a concise overview of the 27 novel fixing strategies identified in our study. For each fix, the table reports: he associated DASP vulnerability category, a short description of the strategy, the number of supporting commits observed in our dataset, and the average expert score obtained from the questionnaire-based evaluation (calculated as the mean of generalizability, sustainability, and effectiveness ratings; see Section 7.1).

This summary allows readers to quickly grasp the diversity of the proposed strategies, compare their empirical support with their perceived quality and relevance, and appreciate at a glance how novel solutions are distributed across different vulnerability classes. While the detailed qualitative discussion of each fix is provided earlier in the text, the tabular overview offers a compact reference that complements the in-depth analysis.

**Summary of Findings for $RQ_2$**

From the analysis of 143 commits not aligned with literature guidelines, we identified 27 novel fixing strategies across DASP categories. Common themes include context-specific solutions (notably for *Time Manipulation* and *Denial of Service*), pragmatic practices such as state tracking, deposit or temporal thresholds, and stricter input validation. Fixes for categories like *Bad Randomness* and *Short Address* focused on removing attacker-controlled sources of unpredictability or avoiding unsafe assumptions. Overall, these results show that developers introduce creative, situation-driven strategies when no clear guidelines exist, highlighting opportunities to formalize new best practices.

## 7 Evaluation of New Fixes

To assess the quality and long-term reliability of the proposed fixes for smart contract vulnerabilities, we conducted a two-pronged empirical evaluation. The first relied on a structured questionnaire to gather expert feedback on three key dimensions: generalizability, long-term sustainability, and effectiveness of each fix. The second involved mining and analyzing historical commit data from real-world repositories to investigate how fixes persist, evolve, or are revised over time. Together, these complementary approaches provide a comprehensive view of the practical impact and robustness of the correction strategies proposed in this study.

**Table 6** Summary of 27 novel fixing strategies

| Category | Fix ID | Description | # Commits | Avg. Expert Score |
|---|---|---|---|---|
| Access Control | Fix 1 | Refund to explicit address instead of `msg.sender` | 1 | 3.85 |
| | Fix 2 | Add whitelist modifier with mapping(address → bool) | 1 | 4.07 |
| Arithmetic | Fix 1 | Limit addend or subtrahend to avoid arithmetic | 3 | 3.52 |
| | Fix 2 | Custom `subMax0` returns 0 on error | 1 | 3.41 |
| | Fix 3 | Save balance before swap to prevent underflow | 1 | 3.89 |
| | Fix 4 | Add boundary check in array loop | 1 | 4.00 |
| | Fix 5 | Cast to `uint96` before shift | 1 | 3.70 |
| | Fix 6 | Scale unit vars to prevent overflow | 2 | 3.96 |
| | Fix 7 | Check buffer before decrement | 2 | 3.96 |
| | Fix 8 | Ignore bids >`uint240.max` instead of reverting | 1 | 4.04 |
| Bad Randomness | Fix 1 | Block multiple seed choices by attacker | 2 | 3.82 |
| | Fix 2 | Remove `msg.sender` from randomness input | 1 | 3.74 |
| Denial of Service | Fix 1 | Track state to prevent redundant actions | 1 | 3.71 |
| | Fix 2 | Set loop upper bound to limit gas use | 1 | 3.67 |
| | Fix 3 | Daily borrow limits block abuse | 1 | 3.82 |
| | Fix 4 | Require min deposit to deter spam | 1 | 2.96 |
| Front Running | Fix 1 | Add `tx.origin` to salt for unpredictability | 1 | 3.85 |
| | Fix 2 | Place `initializer` before `onlyGovernance` | 1 | 3.82 |
| Reentrancy | Fix 1 | Update state before external calls | 2 | 3.96 |
| | Fix 2 | Move `beforeTokenTransfer` after state updates | 1 | 4.04 |
| | Fix 3 | Require NFT balance unchanged after transfer | 1 | 3.93 |
| | Fix 4 | Use `call()` instead of `transfer()/send()` | 1 | 3.82 |
| | Fix 5 | Restrict reentrant functions to internal use | 1 | 3.82 |
| Short Address | Fix 1 | Check `msg.data.length` with modifier | 1 | 4.04 |
| Time Manipulation | Fix 1 | Use input param instead of `block.timestamp` | 1 | 3.21 |
| | Fix 2 | Replace direct timestamp check with reward interval | 1 | 3.52 |
| Unchecked Return Values | Fix 1 | Use OpenZeppelin SafeTransfer to wrap ERC20 calls | 3 | 4.44 |

### 7.1 Expert Feedback: Results and Interpretation

We received a total of nine responses: five from academics and researchers and four from industry professionals, particularly from ICT firms. We specifically sought participants with expertise in both decentralized application (dApp) development and smart contract security.

Regarding professional experience, eight respondents reported between five and eight years of Solidity development, while one respondent reported over ten years of experience. As for familiarity with the DASP TOP 10 taxonomy, five participants declared themselves well-acquainted with it, three had heard of it but were not familiar with its specifics, and one was completely unfamiliar with it.

The collected ratings are summarized in Table 7. Each fix is evaluated across the three dimensions, with statistical indicators such as mean, standard deviation, variance, and mode. From a statistical perspective, the analysis highlights notable trends across categories. Fixes addressing vulnerabilities such as *Reentrancy* and *Unchecked Return Values*

**Table 7** Statistical analysis of generalizability, long-term sustainability and effectiveness of new fixes

| Category | Generalizability | | | | Long-term Sustainability | | | | Effectiveness | | | |
|---|---|---|---|---|---|---|---|---|---|---|---|---|
| | Mean | Std Dev | Variance | Mode | Mean | Std Dev | Variance | Mode | Mean | Std Dev | Variance | Mode |
| Access Control | | | | | | | | | | | | |
| Fix 1 | 3.56 | 1.24 | 1.53 | 4 | 3.67 | 0.71 | 0.50 | 3 | 4.33 | 0.71 | 0.50 | 5 |
| Fix 2 | 4.11 | 1.05 | 1.11 | 5 | 3.67 | 0.71 | 0.50 | 4 | 4.44 | 0.73 | 0.53 | 5 |
| Arithmetic | | | | | | | | | | | | |
| Fix 1 | 3.44 | 1.33 | 1.78 | 2 | 3.33 | 1.22 | 1.50 | 2 | 3.78 | 1.30 | 1.69 | 5 |
| Fix 2 | 3.22 | 1.09 | 1.19 | 4 | 3.33 | 0.87 | 0.75 | 3 | 3.67 | 0.87 | 0.75 | 3 |
| Fix 3 | 3.78 | 1.09 | 1.19 | 3 | 4.00 | 0.87 | 0.75 | 3 | 3.89 | 0.93 | 0.86 | 3 |
| Fix 4 | 4.00 | 1.00 | 1.00 | 4 | 3.78 | 0.97 | 0.94 | 4 | 4.22 | 0.67 | 0.44 | 4 |
| Fix 5 | 3.56 | 1.33 | 1.78 | 3 | 3.33 | 0.87 | 0.75 | 3 | 4.22 | 1.36 | 1.86 | 5 |
| Fix 6 | 3.89 | 1.36 | 1.86 | 5 | 3.78 | 0.97 | 0.94 | 4 | 4.22 | 0.83 | 0.69 | 5 |
| Fix 7 | 4.00 | 1.32 | 1.75 | 5 | 3.78 | 1.20 | 1.44 | 4 | 4.11 | 1.17 | 1.36 | 5 |
| Fix 8 | 3.89 | 1.45 | 2.11 | 5 | 4.00 | 1.12 | 1.25 | 5 | 4.22 | 1.09 | 1.19 | 5 |
| Bad Randomness | | | | | | | | | | | | |
| Fix 1 | 3.89 | 1.36 | 1.86 | 5 | 3.67 | 1.32 | 1.75 | 3 | 3.89 | 1.45 | 2.11 | 5 |
| Fix 2 | 3.56 | 1.51 | 2.28 | 5 | 3.89 | 1.54 | 2.36 | 5 | 3.78 | 1.56 | 2.44 | 5 |
| Denial of Service | | | | | | | | | | | | |
| Fix 1 | 3.56 | 1.24 | 1.53 | 4 | 3.67 | 1.22 | 1.50 | 4 | 3.89 | 1.36 | 1.86 | 5 |
| Fix 2 | 3.56 | 1.01 | 1.03 | 3 | 3.44 | 1.01 | 1.03 | 3 | 4.00 | 1.12 | 1.25 | 5 |
| Fix 3 | 3.67 | 1.12 | 1.25 | 4 | 3.89 | 1.36 | 1.86 | 5 | 3.89 | 1.27 | 1.61 | 4 |
| Fix 4 | 2.89 | 1.27 | 1.61 | 2 | 2.89 | 1.45 | 2.11 | 3 | 3.11 | 1.17 | 1.36 | 3 |
| Front-Running | | | | | | | | | | | | |
| Fix 1 | 3.78 | 1.30 | 1.69 | 4 | 3.67 | 1.32 | 1.75 | 4 | 4.11 | 1.05 | 1.11 | 5 |
| Fix 2 | 3.67 | 1.50 | 2.25 | 5 | 3.78 | 1.48 | 2.19 | 5 | 4.00 | 1.32 | 1.75 | 5 |
| Reentrancy | | | | | | | | | | | | |
| Fix 1 | 4.00 | 1.32 | 1.75 | 5 | 3.89 | 1.36 | 1.86 | 5 | 4.00 | 1.12 | 1.25 | 5 |
| Fix 2 | 4.00 | 1.22 | 1.50 | 5 | 4.00 | 0.87 | 0.75 | 3 | 4.11 | 0.93 | 0.86 | 5 |
| Fix 3 | 3.89 | 1.17 | 1.36 | 5 | 3.78 | 1.09 | 1.19 | 3 | 4.11 | 1.05 | 1.11 | 5 |

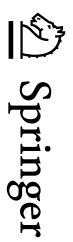

**Table 7** (continued)

| Category | Generalizability | | | | Long-term Sustainability | | | | Effectiveness | | | |
|---|---|---|---|---|---|---|---|---|---|---|---|---|
| | Mean | Std Dev | Variance | Mode | Mean | Std Dev | Variance | Mode | Mean | Std Dev | Variance | Mode |
| Fix 4 | 3.89 | 1.36 | 1.86 | 5 | 3.78 | 1.30 | 1.69 | 5 | 3.78 | 1.39 | 1.94 | 5 |
| Fix 5 | 3.89 | 1.36 | 1.86 | 5 | 3.56 | 1.24 | 1.53 | 4 | 4.00 | 1.41 | 2.00 | 5 |
| Short Address | | | | | | | | | | | | |
| Fix 1 | 3.78 | 1.09 | 1.19 | 5 | 3.89 | 1.05 | 1.11 | 4 | 4.44 | 0.73 | 0.53 | 5 |
| Time Manipulation | | | | | | | | | | | | |
| Fix 1 | 3.00 | 1.20 | 1.43 | 3 | 3.13 | 1.25 | 1.55 | 3 | 3.50 | 1.41 | 2.00 | 4 |
| Fix 2 | 3.22 | 1.20 | 1.44 | 3 | 3.67 | 1.32 | 1.75 | 3 | 3.67 | 1.32 | 1.75 | 3 |
| Unchecked Return Values for Low Level Call | | | | | | | | | | | | |
| Fix 1 | 4.33 | 0.87 | 0.75 | 5 | 4.56 | 0.73 | 0.53 | 5 | 4.44 | 1.01 | 1.03 | 5 |

*for Low Level Call* consistently received higher mean values across all three dimensions, accompanied by relatively low standard deviation and variance. These metrics suggest not only a strong perceived quality of these fixes but also a high degree of consensus among respondents, reinforcing the idea that such fixes are both effective and stable over time. In contrast, categories like *Time Manipulation* and *Arithmetic* display lower average scores and higher variability, indicating that the perceived quality of the solutions may be highly context-dependent or that the proposed strategies are still immature or less convincing for experienced developers.

The dimension of generalizability is particularly polarized. While *Reentrancy* fixes reach values above 4.0 with low dispersion, indicating high confidence in their adaptability, fixes in the *Time Manipulation* category have both lower means and higher variance, suggesting they may be perceived as more tailored to specific scenarios or lacking broader applicability.

Regarding long-term sustainability, the results reflect a similar distribution. Fixes with high average scores and low variance, such as those for *Unchecked Return Values*, indicate that respondents believe these corrections are structurally sound and maintainable over time. On the other hand, the broader standard deviations observed in categories like *Bad Randomness* and *Denial of Service* may reflect uncertainty about how these fixes will behave under evolving operational conditions or in more complex systems.

In terms of effectiveness, the majority of fixes achieved a mode of 5, denoting that most respondents considered them highly effective. However, this unanimity is sometimes contradicted by substantial standard deviation values, particularly in the *Arithmetic* and *Denial of Service* categories. This suggests divergent opinions, possibly due to varied experiences or differing interpretations of what constitutes effectiveness in practice. For instance, a fix may theoretically eliminate a vulnerability but may introduce performance overheads or reduce modularity, influencing subjective assessments.

To deepen the statistical interpretation, we examined the distribution of scores across the three dimensions using a boxplot, shown in Fig. 34. The plot displays the median, interquartile range, and outliers, offering insight into the central tendency and variability of perceptions across evaluation dimensions. This visualization clearly shows that *Effectiveness* is the most highly rated and consistent dimension overall, with a median just above 4 and a tight interquartile range. This suggests broad agreement on the impact and practical utility of the fixes. *Generalizability* and *Long-term Sustainability*, although still positively rated, exhibit more dispersion. The whiskers for Generalizability extend from approximately 2.9 to 4.3, indicating that while many fixes are seen as broadly applicable, some are considered highly context-specific. Sustainability shows a similar pattern but with slightly less spread, reflecting moderate consensus about the maintainability of most solutions. These distributions help clarify not only how the fixes perform on average, but also how consistently they are perceived across different evaluators. The presence of outliers, especially in Generalizability, further underscores the importance of tailoring certain fixes to specific use cases.

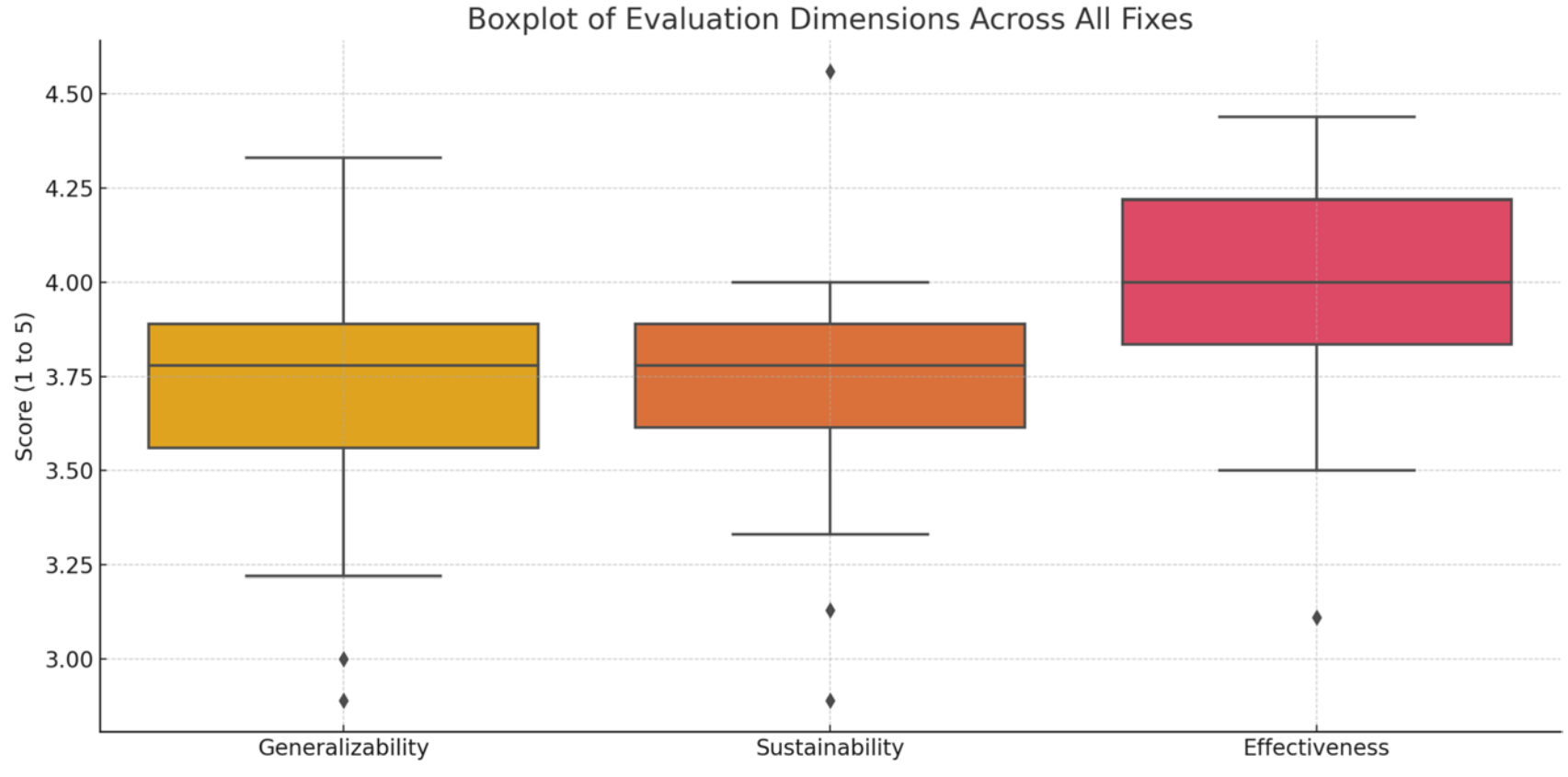

**Fig. 34** Boxplot of respondent scores for Generalizability, Long-term Sustainability, and Effectiveness across all fixes

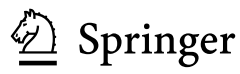

Overall, this statistical analysis enables a nuanced assessment of the proposed correction strategies. The aggregated data clearly indicate that some fixes, particularly in the *Reentrancy* and *Unchecked Return Values* categories, are not only considered highly effective but also generalizable and sustainable. In contrast, other categories reflect greater variability, signaling a need for refinement or more context-sensitive implementations.

The analysis of the collected data revealed that the overall evaluations of the new correction strategies provide valuable insights into their applicability and perceived robustness. In particular, the results suggest that some of the new solutions may be suitable for integration into current development practices, with potential benefits for security and maintainability. These findings, combined with the qualitative comments collected during the study, offer an informative picture of the operational relevance and maturity of each fix. They represent a useful step toward the development of more structured and reliable security guidelines for smart contracts. Moreover, they provide a basis for future research that could further investigate the practical deployment of these strategies, their interaction with other security patterns, and their evolution in response to emerging threats.

Qualitative feedback from experts further explained some of the observed variability in scores. Fixes for *Time Manipulation* and *Denial of Service* were often judged as highly context-dependent, while one arithmetic fix that returned 0 instead of raising an error was considered incomplete. Conversely, established patterns such as *Checks-Effects-Interactions* for *Reentrancy* were seen as standard practices, which may have contributed to their consistently high ratings.

These qualitative insights complement the statistical findings by clarifying why certain fixes received lower ratings despite being effective in principle. To further assess the robustness of the proposed strategies, we extended our evaluation beyond expert judgment by analyzing their long-term stability in real-world repositories, as described in the following section.

### 7.2 Post-Fix Evolution Analysis through Repository Mining

Starting from the set of fixing commits previously collected and filtered during $\mathbf{RQ}_1$, we extracted all subsequent commits that modified the same files involved in the fixes. For each fixing commit, we identified the associated file and traversed the commit history of the corresponding repository to gather all later commits that performed further modifications to those files. We excluded merge commits and retained only those that introduced actual changes. This procedure allowed us to build a dataset of later commits, which contains modifications related to the previously fixed vulnerable code. The extraction was performed using PyDriller, and for each relevant commit, we collected its metadata (e.g., hash, author, date, and message), the corresponding code diff, and the post-commit version of the modified file. The size of such a dataset is 6716 records.

Starting from the dataset of later commits, we aimed to identify cases where the same types of vulnerabilities might have reoccurred after the initial fix. To achieve this, we developed a script that analyzes subsequent commits to the fixing ones we collected that could represent post-fix security patches. In particular, we leveraged NLP techniques, relying even in this case on Spacy to detect commits whose messages contain security-related terminology (e.g., *fix*, *security*, *vulnerability*).

For each project, we searched for commits containing these keywords and associated them with their previously identified fixing commit. We excluded commits already marked as relevant in the initial dataset. This process allowed us to collect a set of 10 security fixes that appeared after the initial fixing commits. One of the authors manually inspected these commits, and all but one preserved the original fix introduced in the earlier commit. Eight of these commits were performed after a fixing commit that adhered to literature guidelines, including the one that did not preserve the fix previously done.

Relying only on the commit message may result in letting pass out important details, to add more depth to the post-fix analysis we systematically evaluated the modification of the later commits. In detail, for each commit identified in the set produced during $\mathbf{RQ}_1$, we extracted the lines added to the fixed file and checked whether all of these lines were still present in the source code of the later commits. If at least one of these lines was missing, we marked the commit for subsequent manual inspection. 201 commits were further evaluated in this way, each diff was viewed on the GitHub page of the associated repository. Hence, one author categorized the modifications, as Table 8 reports. Thus, we describe the category used in the classification task:

- Changes in the business logic: Changes in the functional requirements met by the source code;
- General refactoring: General refactoring operations;
- Gas optimization: Gas optimizations;
- Whitespace, comment or message difference: Modifications in the file that do not change or introduce new logic content;
- Extract variable or method: Extraction of variable, variable declaration, and initialization before using it, as well as for methods;
- Replace a general type with a specific type: Changing a given type or access modifier with a more specific one, for instance, `uint ->uint128`;
- Improved fix: Improvement of the fixes;

**Table 8** Distribution of motivations

| Motivation | Occurrences |
|---|---|
| Changes in business logic | 70 |
| General refactoring | 57 |
| Gas optimization | 23 |
| Whitespace, comment, or message difference | 21 |
| Extract variable or method | 19 |
| Replace a general type with a specific type | 4 |
| Improved fix | 4 |
| Replace a specific type with a general type | 3 |

– Replace a specific type with a general type: Changing a specific type or access modifier with a more general one, for instance, `uint128 ->uint`;

None of the four fix improvements was performed on a fix in our set of new fixing approaches. Overall, this analysis confirms the stability of the collected fixing commits over time. Moreover, we can conclude that fixes are definitive and generally involve modification in a few lines of code, as already discussed by Zhou et al. (2023b).

# 8 Discussion

This section discusses the main findings of the study and practical development behaviors.

## 8.1 Results Discussion

The results show that adherence to academic guidelines is low or even null for certain vulnerabilities. This reflected the dedication of current research to some categories of security threats, such as Reentrancy and Arithmetic. The motivations for this are enclosed in the severity and the popularity of these kinds of vulnerabilities. Indeed, previous research showed how Reentrancy and Arithmetic are more diffused than other categories (Durieux et al. 2020).

The spread of such vulnerabilities is reflected in academic research. To provide deep insight into this, we report the number of papers containing “smart contract” or “smart contracts” and “name of the vulnerability” and “vulnerability” or “vulnerabilities” in the title. Paraphrasing of the vulnerability category name was considered. Hence, Table 9 indicates the count of papers responding to the query for each vulnerability category.

To collect such counts, we leveraged SerpAPI[6], a Google scraper that can work with the engine of Google Scholar. Overall, this result reinforces the hypothesis that literature adherence is higher for some categories as these are more studied in the current literature. In addition, in this scenario, we can conclude that the low or null following of literature guidelines may be correlated to the lack of academic studies or fixing strategies for vulnerability classes, such as bad randomness.

**Table 9** Aggregated count of vulnerabilities

| Vulnerability Class | Count |
|---|---|
| Access Control | 3 |
| Arithmetic | 13 |
| Reentrancy | 31 |
| Bad Randomness | 0 |
| Denial of Service | 9 |
| Front Running | 3 |
| Time Manipulation | 0 |
| Short Address | 0 |
| Unchecked Low Level Calls | 10 |

[6] https://serpapi.com/

The access control category should be considered in a diverse way, due to the presence of this category also in OWASP TOP 10, which reports the most common vulnerability typologies in traditional web apps. Therefore, such a kind of threat is already well-known by the developers.

To address the gaps posed by the low adherence for specific vulnerabilities, future research should be devoted to going alongside developers' behaviors, to enrich the available guidelines in an ever-changing world such as blockchain development. This should be done also by periodically reviewing the output of blockchain technology associations, such as Consensys, and specifically Consensys Diligence[7], which is involved in Ethereum policy discussions and security audits for SCs.

The new fixing strategies identified in this study can guide Solidity developers in addressing security threats by providing them with a broader range of options for managing security vulnerabilities. On the other hand, these approaches improve academic guidelines by incorporating patching procedures used in real practice, thus bridging the gap between academic research and developers' methods to mitigate security problems.

Our analysis revealed that in several commits, developers either removed `send/transfer` in favor of `call`, or replaced SafeMath with Solidity 0.8+ built-in checks. This indicates that developers' fixing strategies are evolving in response to language changes. Therefore, the reliance on Solidity updates is not only a theoretical implication but also observed in practical codebases. These points and implications are deeply discussed next.

## 8.2 Reliance on Mitigation based on Solidity Updates and new Features

The Solidity language has often met the predominant vulnerability-addressing requirement. After the DAO attack, it introduces `send()` and `transfer()` functions, which came with a limited amount of gas to prevent state modifications. Using such functions has been reported as a reentrancy fix in many studies (Chen et al. 2020, 2023a; Zhou et al. 2023a). The Ethereum Improvement Proposal (EIP) 1884 raises the gas cost associated with the `SLOAD` operation, which may cause some existing smart contracts to malfunction. These contracts will encounter issues because their fallback functions previously required less than 2300 gas, but they now exceed this limit. Therefore, gas costs can vay in the future.

This underscores possible issues for contracts whose developers are supposed to be reentrancy bullet-proof without using the `call()` function, as well as for each gas-related problem that might occur. Thus, we pinpoint the need to study more deeply the implications of relying solely on `send` and `transfer` functions. This seems to have already been received to some extent by the developer as we found a few commits that involved the removal of these functions in favor of `call()`.

Another point that needs to be stressed is the reliance on the default arithmetic check introduced with Solidity 0.8+. Since transactions that induce overflow and underflow are reverted, gas costs related to this behavior must be considered, and developers must deal with the transaction revert, handling it. Causing an overflow and letting the default check take care of it, results in a revert that is not accompanied by a detailed message, as we show in Fig. 35. This could cause difficulties with error comprehension.

As Solidity, when releasing the arithmetic default check state that:

[7] https://diligence.consensys.io/

```
1  // SPDX-License-Identifier: MIT
2  pragma solidity 0.8.0;
3
4  contract Overflow{
5
6      function overflow(uint8 a, uint8 b) pure  public returns (uint8){    infinite gas
7          return a + b;
8      }
9  }
```

```
0   Listen on all transactions

CALL  [call] from: 0x5B38Da6a701c568545dCfcB03FcB875f56beddC4 to: Overflow.overflow(uint8,uint8) data: 0x56e...000fa

from             0x5B38Da6a701c568545dCfcB03FcB875f56beddC4
to               Overflow.overflow(uint8,uint8) 0xD7ACd2a9FD159E69Bb102A1ca21C9a3e3A5F771B
execution cost   773 gas (Cost only applies when called by a contract)
input            0x56e...000fa
output           0x4e487b710000000000000000000000000000000000000000000000000000000000000011
decoded input    {
                         "uint8 a": 250,
                         "uint8 b": 250
                 }
decoded output   {
                         "0": "uint8: 0"
                 }
logs             []
raw logs         []

call to Overflow.overflow errored: Error occurred: revert.

revert
        The transaction has been reverted to the initial state.
```

**Fig. 35** Example of an overflow error handled by Solidity's default arithmetic checks introduced in version 0.8+, alongside the revert message

> "Checks for overflow are very common, so we made them the default to increase readability of code, even if it comes at a slight increase of gas costs."

Such variation in readability and gas should be studied, considering as a baseline the SafeMath usage, as it is the most used way to address arithmetic issues and reported in work we reviewed as a best practice to go through this (Zhou et al. 2023a).

## 8.3 Utilization of Contract Vulnerability Handling vs. Library-based Vulnerability Handling

Not using external libraries reduces risks associated with vulnerabilities or errors in imported libraries. eliminates the risk that a library might be compromised in the future, and reduces the risk of losing control or understanding over the flow of execution in code. Moreover, it allows for tailored customizations specific to your use case, as the code is not reliant on external codebases, as we found in some commits, for instance, in the one shown in Fig. 10. Importing libraries can reduce deployment gas costs, but may increase execution costs. Calls to an external library, which incur a fee for each call, might end up being more costly than the one-time deployment expenses (Di Sorbo et al. 2022). Indeed, if the checks are optimized, it is possible to reduce gas consumption compared to an external library.

Generic libraries like SafeMath tend to include universal checks that might not be necessary for all contracts. Even though, Kondo et al. found that the SafeMath.sol library is the most commonly reused code block in smart contracts (Kondo et al. 2020). As a result, redundant runtime checks may lead to significant wastes of gas, as well as time and energy (Gao et al. 2021). Misusing library resources can result in contract defects that lead to financial losses. Huang et al. analyzed 1,018 real-world contracts, pinpointing 905 cases of misuse across 456 of these contracts. This finding indicates that library misuse is a common issue (Huang et al. 2024b). They also found that in their sample 25% of libraries were just used in a single contract.

On the other hand, using libraries to keep the contract code readability high, speeds up the development and increases the maintainability of the code, as such libraries are commonly used. This poses fertile ground for studying developers' awareness regarding library usage and investigating the best gas-saving patterns to prevent vulnerabilities.

## 9 Threats to Validity

**Construct Validity** Construct validity threats primarily arise from errors in manually tagging the relevance of each commit and its associated vulnerability class. To address this issue, two evaluators independently tagged each instance and resolved any conflicts through discussion. Furthermore, the manual evaluation resulted in a very high Cohen's kappa value, indicating strong inter-rater reliability. Another minor threat relates to the use of a visualization tool for annotating code differences. Specifically, we used the tool Diff2HTML[8] to render GitHub commit diffs in a user-friendly browser interface. The visualization was generated using a custom function that opens the diff in the browser via a preconfigured URL. While the tool only affects presentation (not the actual content of the diffs), features like word-level highlighting and side-by-side layout may subtly influence how changes are interpreted. We acknowledge this as a minor potential source of bias in the manual annotation process.

**Internal Validity** A potential threat that might influence our results relates to whether each fix is accurately recognized in the existing literature. To mitigate this threat, we conducted double and independent analyses. Similarly, the same approach was applied when determining if a given fix was overlooked by the state of the art. To further minimize bias, we involved three authors in the conflict resolution process for this step.

**External Validity** The sample under study may not fully reflect real-world conditions. Specifically, a contract in our sample might be part of projects hosted on GitHub as open repositories but may not be deployed on the blockchain. Such information is typically not obtainable from GitHub repositories. However, we could assume that contracts in projects with at least ten stars are not toy projects, Thus, we expect that most contracts of our sample are actively deployed on the blockchain. The choice of the DASP taxonomy may limit the generalizability of our findings, as alternative taxonomies could group vulnerabilities differently or include more recent categories. While DASP is still in use in recent literature, its coverage might not fully reflect the evolving landscape of smart contract vulnerabilities.

[8] https://diff2html.xyz

## 10 Conclusion and Future Work

In this paper, we analyzed the content of 364 commits, each representing changes that address Smart Contract security vulnerabilities categorized according to the DASP TOP 10 taxonomy—a widely recognized classification of common issues in the domain (Durieux et al. 2020). Each commit was considered relevant following a double-checked manual evaluation process, including independent labeling and consensus-based conflict resolution.

Our study pursued two main objectives. First, we aimed to measure the degree to which Solidity developers adhere to established vulnerability mitigation guidelines as documented in the literature. Second, we sought to uncover and characterize fixing strategies that, while used in practice, have not yet been systematically captured in academic work. Through this twofold investigation, we identified 27 distinct and actionable correction strategies that expand the current understanding of how security issues are addressed in real-world smart contract development.

Our results show that developers tend to closely follow recommended practices for certain well-studied vulnerability classes—such as Reentrancy and Arithmetic issues—demonstrating a clear alignment with academic guidance. However, in categories that are less represented or less precisely documented in the literature, such as Time Manipulation or Unchecked Return Values, the adherence is noticeably lower. This finding underscores the presence of gaps between academic knowledge and practical development practices, suggesting that developers are actively experimenting with novel solutions to bridge those gaps. Our study contributes to addressing this disconnect by capturing and analyzing these emerging strategies, thereby enriching the field with practical insights that had not been formally systematized before.

To evaluate the impact, stability, and perceived quality of these new fixes, we conducted a two-pronged empirical evaluation. The first involved a structured expert questionnaire aimed at assessing the generalizability, long-term sustainability, and effectiveness of each proposed fix. The responses gathered from nine experienced professionals in academia and industry, revealed that fixes in categories like Reentrancy and Unchecked Return Values for Low Level Call were not only rated highly across all dimensions but also perceived as robust and reusable. Conversely, categories like Arithmetic and Denial of Service exhibited more variability in responses, reflecting diverse opinions and possibly context-dependent effectiveness. A supporting boxplot visualization highlighted that Effectiveness was consistently rated highest across all fixes, whereas generalizability showed greater dispersion, indicating the need for case-specific adaptation in some scenarios. In addition to the quantitative scores, experts provided qualitative feedback that clarifies why some fixes were perceived less favorably. Several respondents noted that certain fixes, particularly those for *Time Manipulation* and *Denial of Service*, are highly context-dependent and therefore difficult to evaluate without broader information about the contracts from which they were extracted. Others emphasized that some strategies, such as the *Checks-Effects-Interactions* pattern for *Reentrancy*, have become de facto standards rather than optional practices, which may explain their consistently high ratings compared to less consolidated approaches. Concerns were also raised about the completeness of specific fixes—for example, one arithmetic strategy that returned 0 instead of raising an error was considered more of a behavioral change than a true vulnerability mitigation. Finally, experts pointed out that some fixes may derive from the same underlying principle (e.g., arithmetic checks) and could potentially be grouped together, and suggested that including severity levels for vulnerabilities would have provided additional context for evaluating generalizability and sustainability. These insights complement the statistical findings, highlighting not only which fixes were rated highly, but also why certain strategies remain controversial or context-sensitive.

The second part of our evaluation examined the evolution of code after the application of a fix. By tracking more than 6,700 subsequent commits to the same files that contained the original security patches, we investigated whether and how the fixed code changed over time. This analysis, supported by automated filtering and manual inspection, revealed that the majority of the fixes were preserved, indicating their long-term stability. In some cases, improvements were introduced without removing the original logic. We also developed a classification scheme for subsequent modifications—including logic changes, refactoring, and optimization—that provided further evidence of how and why smart contract code evolves after an initial fix.

Overall, our findings offer a comprehensive and empirically grounded picture of how smart contract vulnerabilities are addressed in practice. By identifying not only the fixes commonly used in the field but also assessing their reception by experts and persistence in real-world repositories, this study provides both practical value to developers and analytical depth to the academic community.

Future work may extend this research by exploring patterns and techniques used by developers to optimize gas consumption while maintaining security. This is particularly relevant for contracts with frequent library interactions or repetitive security checks. A comparative analysis of SafeMath usage versus the built-in overflow protections in Solidity versions 0.8 and above could yield valuable insights into the trade-offs between gas efficiency, code readability, and developer preferences. Additionally, further study into the balance between library reuse and custom logic could help identify best practices for minimizing both gas costs and security risks. Furthermore, systematically analyzing the contracts deployed on the blockchain or exploring SC repositories to understand whether certain types of vulnerabilities are more widespread than others, and uncovering the reasons behind these differences would be highly valuable and would deepen our understanding of SC security. For the vulnerability classes that we considered in this study, the adherence to literature guidelines varied considerably. Investigating why developers diverge from academic recommendations would be interesting and crucial to better understanding both developer practices and the adequateness and completeness of the fixing approaches currently known in research.

**Author Contributions** Francesco Salzano and Simone Scalabrino contributed to the study conception and design. Material preparation, data collection, and analysis were performed by Francesco Salzano, Lodovica Marchesi, and Cosmo Kevin Antenucci. The first draft of the manuscript was written by Francesco Salzano and Lodovica Marchesi and all authors commented on previous versions of the manuscript. All authors read, reviewed, and approved the final manuscript. Simone Scalabrino, Rocco Oliveto, and Remo Pareschi supervised the work.

**Funding** Open access funding provided by Università degli Studi del Molise within the CRUI-CARE Agreement. This work is funded by PRIN Project Trust Machines for TrustlessNess (TruMaN): The Impact of Distributed Trust on the Configuration of Blockchain Ecosystems (Identifier Code 2022F5CLN2– CUP H53D23002400006) financed by the Italian Ministry of University and Research, by the National Recovery and Resilience Plan (NRRP), Mission 4 Component 2 Investment 1.5 - Call for tender No.3277 published on December 30, 2021 by the Italian Ministry of University and Research (MUR) funded by the European Union – NextGenerationEU. Project Code ECS0000038 – Project Title eINS Ecosystem of Innovation for Next Generation Sardinia – CUP F53C22000430001 - Grant Assignment Decree No. 1056 adopted on June 23, 2022 by the Italian Ministry of Ministry of University and Research (MUR), and by Università degli Studi di

Cagliari - Prog n. FTE0000522 – CUP: B27H21009670008 - COR: 22390257, sulle risorse di cui all'articolo 3, comma 1, del decreto 24 giugno 2022 per progetto di ricerca e sviluppo MASSIVE – Multi-platform Application to the Self Sovereign Identity Validation Environment, and financed by the Italian Ministry of University and Research, project SOP (Securing sOftware Platforms - CUP: H73C22000890001), as part of the SERICS project (Security and Rights in CyberSpace - n. PE00000014 - CUP: B43C22000750006), and the W.E. B.E.S.T. (Wine EVOO Blockchain Et Smart ContracT) PRIN 2020 project, financed by the Italian Ministry of University and Research (MUR), CUP: F73C22000430001, and this study was partially funded through the NPRR project METROFOOD\-IT. METROFOOD\-IT has received funding from the European Union---NextGenerationEU, NPRR---Mission 4 ``Education and Research'' Component 2: from research to business, Investment 3.1: Fund for the realisation of an integrated system of research and innovation infrastructures---IR0000033 (D.M. Prot. n.120 del 21 June 2022)..

**Data Availibility Statement** The datasets generated and analyzed during the current study are available in the replication package of the study, available at: https://zenodo.org/records/17105939.

## Declarations

**Ethical approval** Not applicable.

**Informed Consent** Not applicable.

**Conflict of Interests** The authors declare no conflict of interest.

## Authors and Affiliations

**Francesco Salzano**[1] · **Lodovica Marchesi**[2] · **Cosmo Kevin Antenucci**[1] · **Simone Scalabrino**[1] · **Roberto Tonelli**[2] · **Rocco Oliveto**[1] · **Remo Pareschi**[1]

✉ Francesco Salzano
francesco.salzano@unimol.it

Lodovica Marchesi
lodovica.marchesi@unica.it

Cosmo Kevin Antenucci
c.antenucci2@studenti.unimol.it

Simone Scalabrino
simone.scalabrino@unimol.it

Roberto Tonelli
roberto.tonelli@unica.it

Rocco Oliveto
rocco.oliveto@unimol.it

Remo Pareschi
remo.pareschi@unimol.it

1 University of Molise, Pesche, Italy

2 University of Cagliari, Cagliari, Italy